\documentclass[twocolumn]{aastex63}

\newcommand  \Hubble   {\ifmmode {\rm km\,s}^{-1}\,{\rm Mpc}^{-1}
                        \else km\,s$^{-1}$\,Mpc$^{-1}$\fi}
\newcommand  \Msun     {\ifmmode M_{\odot} \else $M_{\odot}$\fi}
\newcommand  \Lsun     {\ifmmode L_{\odot} \else $L_{\odot}$\fi}
\newcommand  \cms      {\ifmmode {\rm cm\,s}^{-1} \else cm\,s$^{-1}$\fi}
\newcommand  \acc     {\ifmmode {\rm km\,s}^{-2} \else km\,s$^{-2}$\fi}
\newcommand  \kms      {\ifmmode {\rm km\,s}^{-1} \else km\,s$^{-1}$\fi}

\newcommand  \ergs     {\ifmmode {\rm ergs\,s}^{-1} \else ergs s$^{-1}$\fi}
\newcommand  \ergcms   {\ifmmode {\rm ergs\,cm}^{-2}\,{\rm s}^{-1}
                        \else ergs\,cm$^{-2}$\,s$^{-1}$\fi}
\newcommand  \ergcmsA  {\ifmmode{\rm ergs\,cm}^{-2}\,{\rm s}^{-1}\,{\rm\AA}^{-1}
                        \else ergs\,cm$^{-2}$\,s$^{-1}$\,\AA$^{-1}$\fi}
\newcommand  \ergcmsHz {\ifmmode{\rm ergs\,cm}^{-2}\,{\rm s}^{-1}\,{\rm Hz}^{-1}
                        \else ergs\,cm$^{-2}$\,s$^{-1}$\,Hz$^{-1}$\fi}
\newcommand  \phcms    {\ifmmode {\rm photons\,cm}^{-2}\,{\rm s}^{-1}
                        \else photons\,cm$^{-2}$\,s$^{-1}$\fi}
\newcommand  \phcmsA   {\ifmmode {\rm photons\,cm}^{-2}\,{\rm 
                        s}^{-1}\,{\rm\AA}^{-1}
                        \else photons\,cm$^{-2}$\,s$^{-1}$\,\AA$^{-1}$\fi}

\accepted{by the Astrophysical Journal}

\shorttitle{TWO-DECADE REVERBERATION MAPPING STUDY OF HIGH-LUMINOSITY QUASARS}

\shortauthors{KASPI ET AL.}

\begin{document}

\title{Taking a Long Look: \\ A Two-Decade Reverberation Mapping Study of High-Luminosity Quasars}

\correspondingauthor{Shai Kaspi}
\email{shaik@tauex.tau.ac.il}

\author[0000-0002-9925-534X]{Shai Kaspi} 
\affiliation{School of Physics and Astronomy and Wise Observatory,
Raymond and Beverly Sackler Faculty of Exact Sciences, 
Tel-Aviv University, Tel-Aviv 69978, Israel}

\author[0000-0002-0167-2453]{W. N. Brandt} 
\affiliation{Department of Astronomy and Astrophysics, 525 Davey
Laboratory, Pennsylvania State University, University Park, PA 16802, USA}
\affiliation{Institute for Gravitation and the Cosmos, The 
Pennsylvania State University, University Park, PA 16802, USA}
\affiliation{Department of Physics, The Pennsylvania State University, 
University Park, PA 16802, USA}

\author[0000-0002-6579-0483]{Dan Maoz}
\affiliation{School of Physics and Astronomy and Wise Observatory,
Raymond and Beverly Sackler Faculty of Exact Sciences, 
Tel-Aviv University, Tel-Aviv 69978, Israel}

\author[0000-0002-6766-0260]{Hagai Netzer} 
\affiliation{School of Physics and Astronomy and Wise Observatory,
Raymond and Beverly Sackler Faculty of Exact Sciences, 
Tel-Aviv University, Tel-Aviv 69978, Israel}

\author[0000-0001-7240-7449]{Donald P. Schneider}
\affiliation{Department of Astronomy and Astrophysics, 525 Davey
Laboratory, Pennsylvania State University, University Park, PA 16802, USA}
\affiliation{Institute for Gravitation and the Cosmos, The 
Pennsylvania State University, University Park, PA 16802, USA}

\author[0000-0003-4327-1460]{Ohad Shemmer}
\affiliation{Department of Physics, University of North Texas, 
Denton, TX 76203, USA}

\author[0000-0001-9920-6057]{C. J. Grier}
\affiliation{Department of Astronomy and Astrophysics, 525 Davey
Laboratory, Pennsylvania State University, University Park, PA 16802, USA}
\affiliation{Institute for Gravitation and the Cosmos, The 
Pennsylvania State University, University Park, PA 16802, USA}
\affiliation{Steward Observatory, The University of Arizona, 933 North
Cherry Avenue, Tucson, AZ 85721, USA}



\begin{abstract}

Reverberation mapping (RM) of active galactic nuclei (AGNs) has been used over
the past three decades to determine AGN broad-line region (BLR) sizes and
central black-hole masses, and their relations with the 
AGN's luminosity. Until recently the sample of objects with RM data
was limited to low-luminosity AGNs ($L_{\rm opt} \lesssim 10^{46}$ \ergs )
and low redshifts ($z \lesssim 0.5$). Here we present results from a
reverberation-mapping project of some of the most luminous and highest
redshift quasars that have been mapped to date. The study is based on almost
twenty years of photometric monitoring of 11 quasars, six of which were
monitored spectrophotometrically for 13 years. This is the longest 
reverberation-mapping project carried out so far on this type of AGNs.
We successfully measure a time lag between the \ion{C}{4}$\,\lambda$\,1549
broad emission line and the quasar continuum in three objects, and measure a
\ion{C}{3}]$\,\lambda$\,1909 lag in one quasar.
Together with recently published data on \ion{C}{4} reverberation mapping, the
BLR size is found to scale as the square root of the UV luminosity over eight
orders of magnitude in AGN luminosity. There is a significant scatter in the
relation, part of which may be intrinsic to the AGNs. Although the \ion{C}{4}
line is probably less well suited than Balmer lines for determination of the
mass of the black hole, virial masses are tentatively computed and in spite
of a large scatter we find that the mass of the black hole scales as the
square root of the UV luminosity.

\end{abstract}




\section{Introduction} \label{sec:intro}

Over the past three decades the reverberation-mapping technique has been used
to determine the size of the broad line region (BLR) in active galactic nuclei
(AGNs) and to infer their black-hole (BH) masses in about 200 objects (for
reviews see Peterson 1993, 2006; Netzer \& Peterson 1997, and references therein).
Reverberation mapping uses the response of the emission lines from the BLR
to continuum variations and measures the time lag of the response,
which is interpreted as an estimate of the light-travel time, and hence
the physical separation, between the BH and the bulk of the BLR gas.
A recent compilation of reverberation-mapped objects 
can be found in Bentz \& Katz (2015)\footnote{See the website 
http://www.astro.gsu.edu/AGNmass/ }, Du et al. (2015, 2016, 2018),
Shen et al. (2016), and Grier et al. (2017, 2019). 

Almost all reverberation-mapping studies have focused on low-redshift
AGNs with optical luminosities
($\lambda L_\lambda (5100 {\rm \AA} )$) lower than $10^{46.5} \ \ergs $.
Reverberation mapping of high-luminosity AGNs is quite challenging;
their BLR sizes are expected to be large and their continuum variations are slow.
Combining these features with the cosmological time dilation at high redshift, the
reverberation mapping campaigns must span a period of a decade or more,
necessitating very long-term commitment of observing facilities. 
In addition, large telescope collecting areas are required due to the
faintness of such quasars. 

Several early reverberation-mapping efforts for high-luminosity quasars
were unsuccessful at detecting reverberation time lags, e.g., Welsh et al.
(2000), Trevese et al. (2006, 2007) and A. Marconi (2005, private
communication), due to the insufficient duration of these projects.
Trevese et al. (2014) reported 11 years of monitoring of the $z=2.048$
quasar PG\,1247+267; however, their sampling was poor (one to three points
a year) and the model-independent methods used to derive the time lag are not
sufficiently accurate: using the model-dependent\footnote{In that it adopts a
reasonable Damped Random Walk based interpolation approach, see Yu et al.
(2019) and Li et al. (2019).} method of Stochastic Process Estimation for
AGN Reverberation (SPEAR) developed by Zu et al. (2011) they claimed a
significant time-lag measurement which, however, we consider questionable.
Saturni et al. (2016) reported a reverberation mapping study of 
APM\,08279+5255 over 12 years, also with a rather limited sampling rate. 
They use the model-dependent JAVELIN method (Zu et al. 2013) a development 
of SPEAR, to estimate the time lag. Inspection of their figure~4 shows that 
the use of JAVELIN introduces several model points inside long gaps in the 
data, i.e., there are no real overlapping measurements in the data used to 
determine the lag. This greatly lowers the significance of this result and we 
will not consider it further here.

Shen et al. (2016) have reported preliminary results of 15 BLR lag measurements 
from the Sloan Digital Sky Survey Reverberation Mapping (SDSS-RM) project. The 
project targets almost 850 AGNs with redshifts up to $z=4.5$ and expects to 
deliver BLR lag measurements for a few hundred. The reported preliminary lag 
measurements are for intermediate-luminosity quasars at $0.3\lesssim z < 0.8$ 
and include nine H$\beta$ lags and six \ion{Mg}{2} lags. 
Further BLR lag measurements from that project have been reported by Grier et al.
(2017) who measured H$\beta$ and H$\alpha$ BLR lag measurements for a total 
of 44 and 18 quasars, respectively, at the redshift range of $0.1 < z < 1.1$.
They find that, in most objects, the time 
delay of the H$\alpha$ emission is consistent with or slightly longer than 
that of H$\beta$. Their black-hole mass measurements are mostly consistent with 
expectations from single-epoch black-hole mass measurements and the local 
$M_{BH}$--$\sigma_{*}$ relation.

Lira et al. (2018) describe a reverberation-mapping project which targeted 17 
high-luminosity quasars for more than 10 years. These authors are able to 
measure statistically significant ($>1\sigma$) time lags for the
emission lines of Ly$\alpha$ (8 objects), \ion{C}{4} (8 objects), 
\ion{Si}{4} (3 objects), and \ion{C}{3}] (1 object). Altogether significant
time lags were measured for 10 distinct objects. Lira et al. (2018)
give an updated \ion{C}{4} radius--luminosity relation, and presented for the
first time a radius--luminosity relation for the other three lines.
They report that the regions responsible for the emission of Ly$\alpha$, 
\ion{Si}{4}, \ion{C}{4}, and \ion{C}{3}] are commonly interior to that 
producing H$\beta$, but there is no clear stratification among them. 
Approximately 18\% (3/17) of their sources show an unexpected behavior in some 
emission lines in which the line light curves do not appear to follow the 
observed UV continuum variations. This is similar to the recent behavior
detected in the AGN NGC\,5548 in which during a monitoring of 180 days the BLR
had a period of $\sim 40$ days in which it did not follow the UV continuum 
variations (Goad et al. 2019).

Hoormann et al. (2019) report \ion{C}{4} reverberation mapping time 
lags for two objects from the Dark Energy Survey Supernova Program (DES-SN)
and the Australian Dark Energy Survey (OzDES) at redshifts of 1.9 
and 2.6. In that study the photometric monitoring covers five years while 
the spectroscopic monitoring was 3--4 years. 
Out of the 393 objects with \ion{C}{4} in their sample, they identified a
subset of objects which were expected to have lags of the order of 1 yr. 
They further cut the sample to 23 objects which were variable and had 
high-cadence data, but only two of them had significant time-lag measurements.

In a recent paper Grier et al. (2019) report on \ion{C}{4} reverberation-mapping
time lags from the Sloan Digital Sky Survey Reverberation Mapping project.
They report time lags for 52 AGNs, with an estimated false-positive detection 
rate of 10\%. 16 of these AGNs are defined as their ``gold sample'' with 
the highest-confidence lag measurements. These 16 AGNs lay in the redshift 
range of $1.4 < z < 2.8$ and luminosity range of 
$10^{44.5} < \lambda L_\lambda (1350 {\rm \AA} ) < 10^{45.6}$ erg\,s$^{-1}$.
Adding these 16 objects to the objects with \ion{C}{4} time lag measurements 
of previous studies they find the radius---luminosity relation has a slope 
of $0.52\pm0.04$. Shen et al. (2019) show how adding photometric data to the
spectroscopic monitoring of this project can improve the detection of time lags. 
They report on three more objects with detected \ion{C}{4} time lags.

In Kaspi et al. (2007; hereafter Paper I) we presented the first results
from our reverberation-mapping campaign of several high-luminosity AGNs.
In this paper we report the final results from this campaign after 18 years
of photometric and 13 years of spectroscopic monitoring. 
This is the longest single reverberation-mapping campaign carried out to date
(only one object surpasses this, the multiple campaigns on NGC\,5548 over
the past two decades).
Section~2 describes the observations and the data-reduction process.
Section~3 presents the results. 
Section~4 analyzes the new results, combined with the results of other
reverberation-mapping campaigns of high-luminosity AGNs that have yielded
significant results.

Throughout this paper we use the standard cosmology with
$H_{0}=70$\,km\,s$^{-1}$\,Mpc$^{-1}$, $\Omega_{M}=0.3$, and
$\Omega_{\Lambda}=0.7$. 

\section{Sample, observations, and data reduction}

The sample, observations, and data processing are described in detail in
Paper I. For completeness we summarize them here and note 
differences where applicable. 

The photometric sample consists of 11 high-luminosity quasars with observed
magnitude of $V\lesssim 18$ and redshift in the range $2\lesssim z\lesssim 3.4$.
The luminosity range
is $10^{46.9}\la\lambda L_\lambda(1350\,{\rm \AA})\la10^{48.0}$ \ergs, which
translates to 
$10^{46.6}\la\lambda L_\lambda(5100\,{\rm \AA})\la10^{47.8}$ \ergs\
(see Table~\ref{tab:luminosities} for details). Six of the
objects in our sample are radio loud and five are radio quiet (see Paper I
for details). All objects are at high declination ($\delta \ga +60\degr$) to
enable largely uninterrupted yearly coverage from the Northern hemisphere.

Photometric observations were obtained at the Wise observatory 1\,m telescope
in the Johnson-Cousins $B$ and $R$ bands, from 1995 to mid-2013. The only
difference from the description of Paper I is that during 2006 the Tektronix
CCD was replaced with a Princeton Instruments camera with a E2V 
$1300\times 1340$ CCD which had similar field of view and quantum efficiency;
exposure time and sampling rate were unchanged from the description in Paper I.
Exposure times for all objects were 250 and 300 sec in the $R$ and $B$ bands,
respectively, except that for S5\,2017+744 we used 300 and 400 sec.
Each object was observed photometrically about once a month for eight months
each year. Photometric uncertainties are of order 0.03 mag.
The data have been reduced using IRAF\footnote{IRAF
(Image Reduction and Analysis Facility) is distributed by the National
Optical Astronomy Observatories, which are operated by AURA, Inc,
under cooperative agreement with the National Science Foundation.}
procedures in the standard way.

Spectroscopic observations were performed for six of the 11 
photometrically-monitored AGNs at the Hobby-Eberly Telescope (HET; Ramsey
et al. 1998). Spectra were obtained starting in 1999 with the Low Resolution
Spectrograph (LRS; Hill et al. 1998), which began its primary observations in
1998, and spectroscopic observations continued until the instrument was 
decommissioned in mid-2013. Spectra were taken 
with a long slit which included the AGN and a field star which was used for 
flux calibration. Observations typically consisted of two consecutive exposures
of the quasar/star pair with exposure times (for each exposure) of 900 sec for
S4\,0636+68 and SBS\,1116+603, 600 sec for S5\,0836+71 and SBS\,1233+594, and
300 sec for SBS\,1233+594 and HS\,1700+6416 (three of these objects are
radio-loud AGNs and three are radio-quiet AGNs). We observed each of the
six objects about three to four times each year with two-to-three month
separation between observations. The total program time amounted to roughly
12 hours of HET time each year (including overheads). The spectroscopic
data were processed using standard IRAF routines as described in Paper~I.

\begin{deluxetable*}{lccr}
\tablecaption{Luminosities  \label{tab:luminosities}}
\tablecolumns{4}
\tablewidth{0pt}
\tablehead{
\colhead{Object}    &
\colhead{$\lambda L_\lambda$(1350 \AA )\tablenotemark{a}} &
\colhead{$\lambda L_\lambda$(5100 \AA )\tablenotemark{a}} &
\colhead{$R$\tablenotemark{b}} }
\startdata
S5 0014+81   & $ 7.66 \pm 1.00$ & 3.94 &   493.3 \\
S5 0153+74   & $ 1.20 \pm 0.20$ & 0.62 & 12377.8 \\
S4 0636+68\tablenotemark{c} & $ 3.66 \pm 0.26$ & 1.88 &   133.2 \\
S5 0836+71\tablenotemark{c}   & $ 1.03 \pm 0.16$ & 0.53 & 10064.5 \\
TB 0933+733  & $ 1.57 \pm 0.18$ & 0.81 &     5.2 \\
SBS 1116+603\tablenotemark{c} & $ 1.79 \pm 0.20$ & 0.92 &   632.7 \\
SBS 1233+594\tablenotemark{c} & $ 1.98 \pm 0.18$ & 1.02 &     1.3 \\
SBS 1425+606\tablenotemark{c} & $ 4.54 \pm 0.37$ & 2.34 &     4.6 \\
HS 1700+6416\tablenotemark{c} & $ 5.00 \pm 0.34$ & 2.57 &     3.6 \\
HS 1946+7658 & $ 10.9 \pm 1.22$ & 5.63 &     1.7 \\
S5 2017+744  & $ 0.72 \pm 0.09$ & 0.37 &  2765.8 \\
\enddata
\tablecomments{
$\lambda L_\lambda$(1350 \AA ) for the six objects in our spectroscopic 
sample were measured from the spectra shown in Figure~\ref{fig:mean_std}, 
for S5\,0014+81, S5\,0153+74, and HS\,1946+7658 from spectra published in the
literature, and for TB\,0933+733 and S5\,2017+74 from extrapolation of their
SED given in NED. Luminosities were corrected for Galactic extinction 
using $A_V$ from NED based on Schlafly \& Finkbeiner (2011).
$\lambda L_\lambda$(5100 \AA ) was computed from $\lambda L_\lambda$(1350 \AA )
using a power law $f_\nu \propto \nu^{-0.5}$. 
}
\tablenotetext{a}{In units of $10^{47}$ erg\,s$^{-1}$.}
\tablenotetext{b}{Radio loudness is the radio (1.4 GHz)-to-optical 
(estimated at 4400 \AA ) flux ratio, from Paper I.}
\tablenotetext{c}{Spectroscopically monitored.}
\end{deluxetable*}

\begin{deluxetable*}{lcccccccc}
\tablecaption{INTEGRATION LIMITS FOR CONTINUUM BANDS AND EMISSION LINES \label{tab:waves}}
\tablecolumns{9}
\tabletypesize{\scriptsize}
\tablewidth{0pt}
\tablehead{
\colhead{Object}    &
\colhead{Ly$\alpha$} &
\colhead{Continuum\tablenotemark{\scriptsize a}} &
\colhead{Continuum} &
\colhead{\ion{C}{4}} &
\colhead{Continuum} &
\colhead{Continuum} &
\colhead{\ion{C}{3}]} &
\colhead{Continuum} }
\colnumbers
\startdata
S4 0636+68   &  4986-5356 & {\bf 5622-5701} & 6071-6141 & 6301-6591 & 6616-6701 & \nodata & \nodata & \nodata \\
S5 0836+71   & \nodata & \nodata & 4581-4676 & 4753-5028 & 5111-5191 & 5769-5852 & 5896-6214 & {\bf 6456-6556} \\
SBS 1116+603 & \nodata & \nodata & 5436-5487 & 5489-5760 & 5854-5939 & 6545-6649 & 6685-7061 & {\bf 7154-7220} \\
SBS 1233+594 & 4538-4867 & 4876-4926 & 5611-5737 & 5750-6004 & {\bf 6065-6165} & \nodata & \nodata & \nodata \\
SBS 1425+606 &  4975-5340 & 5365-5415 & {\bf 6051-6141} & 6261-6641 & 6676-6771 & \nodata & \nodata & \nodata \\
HS 1700+6416 & 4422-4782 & 4792-4855 & 5386-5454 & 5536-5893 & {\bf 5935-5993} & \nodata & \nodata & \nodata \\
\enddata
\tablenotetext{a}{Due to the noisy absorbed spectrum on the blue side of
Ly$\alpha$, only the continuum on the red side was used to define the continuum
underlying this line.}
\tablecomments{Wavelengths are in units of \AA\ in the observer's frame. 
The boldface ranges are the continuum bands used for the cross-correlation analysis.}
\end{deluxetable*}

\begin{figure*}
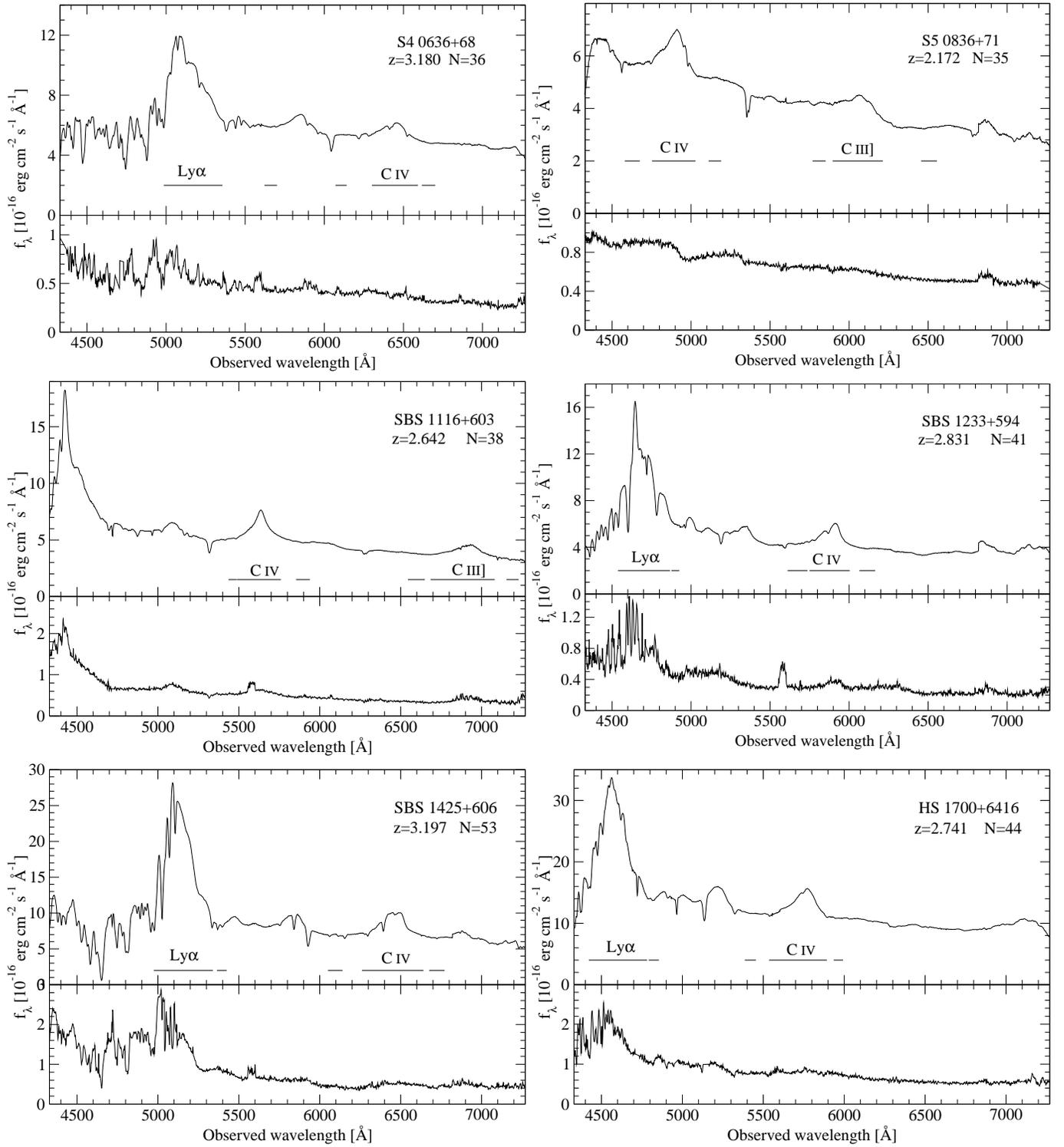

\gridline{\fig{fig1a.eps}{0.5\textwidth}{}
          \fig{fig1b.eps}{0.5\textwidth}{}
          }
\vspace{-2.3cm}
\gridline{\fig{fig1c.eps}{0.5\textwidth}{}
          \fig{fig1d.eps}{0.5\textwidth}{}
          }
\vspace{-2.3cm}
\gridline{\fig{fig1e.eps}{0.5\textwidth}{}
          \fig{fig1f.eps}{0.5\textwidth}{}
          }
\vspace{-2.0cm}
\caption{Mean (top panels) and rms (bottom panels) spectra for the six 
objects in our sample. Under each object name the redshift ($z$) and the 
number of spectra ($N$) used in the mean and rms are given. Vertical 
lines in top panels show the integration ranges for continuum bands 
and emission lines as listed in Table~\ref{tab:waves}. Emission lines 
are labeled and continuum bands have no labels.
\label{fig:mean_std}}
\end{figure*}

Figure~\ref{fig:mean_std} shows the mean and rms\footnote{As defined in 
Equation~1 of Paper~I.} spectra of the six spectroscopically monitored quasars.
We used the mean and rms spectra to identify line-free spectral bands which are
suitable for setting the continuum underlying the emission lines, and the 
wavelength limits for integrating the line fluxes.
The spectral ranges for the lines and the continuum bands on both sides of each
line are given in Table~\ref{tab:waves}. The detailed measurement process is as
described in Kaspi et al. (2000) and in Paper I.

\begin{figure*}
\gridline{\fig{fig2a.eps}{1.0\textwidth}{(a)}
          }
\gridline{\fig{fig2b.eps}{0.5\textwidth}{(b)}
          \fig{fig2c.eps}{0.5\textwidth}{(c)}
          }
\caption{Light curves and CCFs for S\,0636+68. 
(a) Black triangles are photometric data from Wise and red squares are 
spectroscopic data from HET. Data up to the blue vertical dashed line were
published in Paper~I.
(b) and (c) show the CCFs for lines as noted in the figure. The time lag is given 
in the observed frame. ICCF is plotted as the solid line while ZDCF is plotted as 
black points with uncertainties. The blue vertical dashed line denotes
a time lag of zero and the red horizontal dashed line denotes a cross-correlation 
coefficient of zero. No significant peak is identified in either of the CCFs.
\label{fig:s40636}}
\end{figure*}

\begin{figure*}
\gridline{\fig{fig3a.eps}{1.0\textwidth}{(a)}
          }
\gridline{\fig{fig3b.eps}{0.5\textwidth}{(b)}
          \fig{fig3c.eps}{0.5\textwidth}{(c)}
          }
\caption{Light curves and CCFs for S5\,0836+71. Symbols as in Figure~\ref{fig:s40636}.
(b) The peak seen in the \ion{C}{3}] CCF, at $\sim 3600$ days,
is marginally significant. See text for discussion. 
(c) A significant time lag is detected for the \ion{C}{4} line.
\label{fig:s50836} }
\end{figure*}

\begin{figure*}
\gridline{\fig{fig4a.eps}{1.0\textwidth}{(a)}
          }
\gridline{\fig{fig4b.eps}{0.5\textwidth}{(b)}
          \fig{fig4c.eps}{0.5\textwidth}{(c)}
          }
\caption{Light curves and CCFs for SBS\,1116+603. Symbols as in Figure~\ref{fig:s40636}.
(b) and (c) Both CCFs show significant peaks.
\label{fig:sbs1116}}
\end{figure*}

\begin{figure*}
\gridline{\fig{fig5a.eps}{1.0\textwidth}{(a)}
          }
\gridline{\fig{fig5b.eps}{0.5\textwidth}{(b)}
          \fig{fig5c.eps}{0.5\textwidth}{(c)}
          }
\caption{Light curves and CCFs for SBS\,1233+594. Symbols as in Figure~\ref{fig:s40636}.
(b) Although the ICCF for the Ly$\alpha$ line shows a significant peak the 
ZDCF method does not show such a peak. See text for discussion.
(c) No significant time lag is detected for the \ion{C}{4} line.
\label{fig:sbs1233}}
\end{figure*}

\begin{figure*}
\gridline{\fig{fig6a.eps}{1.0\textwidth}{(a)}
          }
\gridline{\fig{fig6b.eps}{0.5\textwidth}{(b)}
          \fig{fig6c.eps}{0.5\textwidth}{(c)}
          }
\caption{Light curves and CCFs for SBS\,1425+606. Symbols as in Figure~\ref{fig:s40636}.
(b) No significant time lag is detcted in the Ly$\alpha$ line.
(c) A significant time lag is detected for the \ion{C}{4} line.
\label{fig:sbs1425}}
\end{figure*}

\begin{figure*}
\gridline{\fig{fig7a.eps}{1.0\textwidth}{(a)}
          }
\gridline{\fig{fig7b.eps}{0.5\textwidth}{(b)}
          \fig{fig7c.eps}{0.5\textwidth}{(c)}
          }
\caption{Light curves and CCFs for HS\,1700+6416. Symbols as in Figure~\ref{fig:s40636}.
(b) and (c) No significant peak is identified in both CCFs.
\label{fig:hs1700}}
\end{figure*}

\begin{deluxetable*}{llrrccrccc}
\tablecolumns{9}
\tabletypesize{\scriptsize}
\tablewidth{0pt}
\tablecaption{Variability Measures
\label{vartab}}
\tablehead{
\colhead{Object}        &
\colhead{Light Curve}   &
\colhead{N\tablenotemark{\scriptsize a}}   &
\colhead{Mean\tablenotemark{\scriptsize b}}     &
\colhead{RMS\tablenotemark{\scriptsize b}} &
\colhead{Mean Uncertainty\tablenotemark{\scriptsize b}} &
\colhead{$\chi^{2}_{\nu}$} &
\colhead{P$(\chi^{2}|{\nu})$\tablenotemark{\scriptsize c}} &
\colhead{$\sigma_N$} &
\colhead{$F_{\rm var}$} \\
\colhead{(1)} &
\colhead{(2)} &
\colhead{(3)} &
\colhead{(4)} &
\colhead{(5)} &
\colhead{(6)} &
\colhead{(7)} &
\colhead{(8)} &
\colhead{(9)} &
\colhead{(10)} 
} 
\startdata
\multicolumn{10}{c}{Spectroscopically Monitored Objects}  \\
\hline   
S4 0636+68 & continuum & 209 &  5.94 &  0.28 &  0.14 & 26.53 & 0    &  4.15 &$0.040\pm0.003$ \\ 
           & Ly$\alpha$&  36 & 12.09 &  0.51 &  0.67 &  1.04 & 0.40 &\nodata&    \nodata     \\ 
           & \ion{C}{4}&  36 &  1.86 &  0.15 &  0.09 &  4.25 & 0    &  4.70 &$0.054\pm0.015$ \\ 
           &     $B$   & 169 & 17.203&  0.061&  0.033&  4.69 & 0    &  4.81 &$0.045\pm0.004$ \\
S5 0836+71 & continuum & 181 &  3.18 &  0.45 &  0.08 &162.13 & 0    & 13.95 &$0.138\pm0.008$ \\
           & \ion{C}{4}&  35 &  2.26 &  0.41 &  0.10 & 14.95 & 0    & 17.41 &$0.173\pm0.022$ \\ 
           &\ion{C}{3}]&  35 &  1.06 &  0.28 &  0.08 & 14.19 & 0    & 25.08 &$0.250\pm0.033$ \\ 
           &     $B$   & 144 & 17.199&  0.178&  0.031& 54.35 & 0    & 16.84 &$0.167\pm0.010$ \\
SBS 1116+603& continuum& 201 &  3.14 &  0.34 &  0.09 & 27.63 & 0    & 10.59 &$0.105\pm0.006$ \\ 
           & \ion{C}{4}&  38 &  3.01 &  0.36 &  0.09 & 19.81 & 0    & 11.53 &$0.115\pm0.014$ \\ 
           &\ion{C}{3}]&  38 &  1.89 &  0.29 &  0.12 &  5.52 & 0    & 13.71 &$0.135\pm0.019$ \\
           &     $B$   & 154 & 17.421&  0.132&  0.034& 20.32 & 0    & 11.60 &$0.115\pm0.007$ \\
SBS 1233+594& continuum& 198 &  4.00 &  0.27 &  0.09 & 33.10 & 0    &  6.30 &$0.062\pm0.004$ \\ 
           & Ly$\alpha$&  41 & 12.62 &  0.84 &  1.26 &  0.56 & 1.00 &\nodata&     \nodata    \\ 
           & \ion{C}{4}&  41 &  2.40 &  0.19 &  0.09 &  5.78 & 0    &  6.87 &$0.067\pm0.010$ \\ 
           &     $B$   & 158 & 17.758&  0.094&  0.037&  7.64 & 0    &  7.89 &$0.077\pm0.005$ \\
SBS 1425+606& continuum& 228 &  6.84 &  0.45 &  0.14 & 28.22 & 0    &  6.24 &$0.061\pm0.003$ \\ 
           & Ly$\alpha$&  53 & 32.99 &  1.48 &  2.01 &  0.41 & 1.00 &\nodata&     \nodata    \\ 
           & \ion{C}{4}&  53 &  5.95 &  0.35 & 0.23& 2.24&$7.7\times10^{-7}$&4.53&$0.044\pm0.008$\\ 
           &     $B$   & 170 & 17.278&  0.102&  0.030& 16.08 & 0    &  8.97 &$0.089\pm0.005$ \\
HS 1700+6416& continuum& 247 & 10.61 &  0.69 &  0.23 & 33.80 & 0    &  6.14 &$0.060\pm0.003$ \\ 
           & Ly$\alpha$&  44 & 30.05 &  1.31 &  2.04 &  0.91 & 0.65 &\nodata&     \nodata    \\  
           & \ion{C}{4}&  45 &  6.86 &  0.41 &  0.24 &  3.46 & 0    &  4.86 &$0.047\pm0.008$ \\ 
           &     $B$   & 197 & 16.068&  0.072&  0.026& 10.83 & 0    &  6.23 &$0.061\pm0.004$ \\
\hline
\multicolumn{10}{c}{Photometrically Monitored Objects}  \\
\hline
S5 0014+81 &     $R$   & 163 & 16.822&  0.021&  0.026& 1.398 &0.00062&\nodata&    \nodata    \\ 
           &     $B$   & 162 & 17.998&  0.050&  0.038& 1.680 &$1.4\times10^{-7}$&3.344&$0.027\pm0.005$ \\
S5 0153+74 &     $R$   & 136 & 17.852&  0.149&  0.040& 26.32 & 0    & 13.27 &$0.131\pm0.009$ \\ 
           &     $B$   & 131 & 18.324&  0.189&  0.065& 24.52 & 0    & 16.11 &$0.154\pm0.012$ \\
TB 0933+733&     $R$   & 163 & 17.334&  0.073&  0.032&  7.83 & 0    &  6.10 &$0.059\pm0.004$ \\ 
           &     $B$   & 159 & 17.607&  0.075&  0.038&  8.80 & 0    &  6.17 &$0.056\pm0.005$ \\
HS 1946+7658&    $R$   & 228 & 16.429&  0.037&  0.025&  3.08 & 0    &  2.60 &$0.025\pm0.002$ \\ 
            &    $B$   & 216 & 17.149&  0.059&  0.031&  4.05 & 0    &  4.64 &$0.045\pm0.003$ \\ 
S5 2017+74 &     $R$   & 189 & 18.718&  0.094&  0.049&  6.29 & 0    &  7.16 &$0.067\pm0.006$ \\
           &     $B$   & 182 & 18.989&  0.110&  0.049& 10.55 & 0    &  9.00 &$0.083\pm0.006$ \\
\enddata
\tablecomments{No data in columns (9) \& (10) indicate that no significant
variability was detected in a light curve. None of
the Ly$\alpha$ light curves show significant variability.}
\tablenotetext{a}{N is the number of points in each light curve. 
The continuum light curve for each object is a merge of the $R$-band light curve
and the continuum spectroscopic light curve. The $R$ band has a few more
points in the light curve than the $B$ band since it is somewhat less sensitive
when observing through bad-weather. Thus, for each object the number of
points in the continuum light curve is slightly larger than the sum of the
number of points in the line light curve and the $B$-band light curve. }
\tablenotetext{b}{Units are
10$^{-16}$~erg\,cm$^{-2}$\,s$^{-1}$\,{\rm \AA}$^{-1}$ for the continuum
light curves, 10$^{-14}$~erg\,cm$^{-2}$\,s$^{-1}$ for the emission-line light
curves, and apparent magnitude for the $R$-band and $B$-band light curves.}
\tablenotetext{c}{Probabilities smaller than $10^{-9}$
are listed as zero.}
\end{deluxetable*}

\section{Results} \label{sec:res}

\subsection{Continuum and Line Light Curves}
\label{subsec:LC}

The light curves for the six spectroscopically monitored objects are presented
in Figures~\ref{fig:s40636} to \ref{fig:hs1700} with the data listed in
Tables~\ref{tab:cont_lc} and \ref{tab:line_lc}. The photometric $R$-band light
curves were merged with the spectroscopic continuum light curves as described
in Paper~I in order to increase the sampling of the continuum light curve.
The different bands used for the flux measurements are listed in 
Table~\ref{vartab} for each object (wavelengths are given in the observed frame).
The number of observations as well as statistical information about the light
curves are provided at the bottom of Table~\ref{vartab}.
Columns are (1) the object name, (2) the particular light curve, (3) the number
of points in that light curve, (4), (5), and (6) the mean ($\bar{f}$), rms
($\sigma$), and the mean uncertainty ($\delta$) of all data points in the light
curves, respectively, in the appropriate units for each light curve, (7) the
$\chi^{2}_{\nu}$ obtained by fitting a constant to the light curve, and (8)
P$(\chi^{2}|{\nu})$, the probability to obtain such $\chi^{2}$ if there were
intrinsically no variability. This test is used to determine whether the light
curve is consistent with a constant flux to a significance level of
95\%. For light curves which show significant variability column
(9) lists the intrinsic normalized variability measure, 
$\sigma_{N} = 100\sqrt{\sigma^{2}-\delta^{2}}/\bar{f}$ 
(used by Kaspi et al. 2000), and (10) is the fractional variability amplitude,
$F_{var}$, and its uncertainty (e.g., Rodriguez-Pascual et al. 1997;
Edelson et al. 2002), which are defined as 
$F_{var}=\sqrt{\sigma^2-<\sigma^{2}_{err}>}/\bar{f}$ and 
$\sigma_{F_{var}}=\sqrt{1/(2N)}\,\sigma^2/(\bar{f}^{2}\,F_{var})$, respectively, 
where $<\sigma^{2}_{err}>$ is the mean error squared.

Further discussion of the photometric light curves as well as the $B$-band
light curves of these objects are given in Appendix~A.

All light curves as well as individual and mean spectra are also available at
https://doi.org/10.5281/zenodo.4718461 and at
the web site: http://wise-obs.tau.ac.il/$\sim$shai/highz/ .

\begin{deluxetable}{cc}
\tablecolumns{2}
\tablewidth{0pt}
\tablecaption{Continuum Light Curves
\label{tab:cont_lc}}
\tablehead{
\colhead{JD}    &
\colhead{$f_\lambda$\tablenotemark{a}}
} 
\startdata
\multicolumn{2}{c}{S4\,0636+68}   \\
\tableline
     2449667.42 & $ 6.340    \pm 0.147 $ \\
     2449728.43 & $ 6.396    \pm 0.172 $ \\
     2449773.24 & $ 6.285    \pm 0.157 $ \\
     2449804.22 & $ 6.285    \pm 0.140 $ \\
     2449954.56 & $ 6.128    \pm 0.193 $ \\
     2449981.58 & $ 6.068    \pm 0.152 $ \\
     2450007.50 & $ 5.672    \pm 0.173 $ \\
     2450035.48 & $ 5.831    \pm 0.167 $ \\
     2450080.23 & $ 5.812    \pm 0.140 $ \\
     2450104.38 & $ 5.864    \pm 0.147 $ \\
\enddata
\footnotesize
\tablecomments{Table~\ref{tab:cont_lc} is presented in its
entirety in the electronic edition of the Astrophysical Journal. A
portion is shown here for guidance regarding its form and content.}
\tablenotetext{a}{In units of 10$^{-16}$~erg\,cm$^{-2}$\,s$^{-1}$\,\AA$^{-1}$.}
\end{deluxetable}

\begin{deluxetable}{cc}
\tablecolumns{2}
\tablewidth{0pt}
\tablecaption{Line Light Curves
\label{tab:line_lc}}
\tablehead{
\colhead{JD}    &
\colhead{$f_\lambda$\tablenotemark{a}}
} 
\startdata
\multicolumn{2}{c}{S4\,0636+68 -- Ly$\alpha$}   \\
\tableline
     2451871.93 & $ 11.553 \pm 0.496 $ \\
     2451935.69 & $ 11.727 \pm 1.093 $ \\
     2452337.66 & $ 11.508 \pm 0.886 $ \\
     2452559.99 & $ 11.790 \pm 0.816 $ \\
     2452618.84 & $ 11.872 \pm 0.722 $ \\
     2452698.65 & $ 12.013 \pm 1.659 $ \\
     2452962.88 & $ 12.577 \pm 0.953 $ \\
     2452970.87 & $ 12.380 \pm 0.656 $ \\
     2453047.68 & $ 13.282 \pm 0.343 $ \\
     2453294.97 & $ 12.770 \pm 0.542 $ \\
\enddata
\footnotesize
\tablecomments{Table~\ref{tab:line_lc} is presented in its
entirety in the electronic edition of the Astrophysical Journal. A
portion is shown here for guidance regarding its form and content.}
\tablenotetext{a}{In units of 10$^{-14}$~erg\,cm$^{-2}$\,s$^{-1}$.}
\end{deluxetable}

\subsection{Time Series Analysis}
\label{sec:time}

In order to measure a time lag that can serve as an estimate for the BLR size
one usually cross correlates the line and continuum light curves. Two commonly 
used methods were applied. The first is the Interpolated Cross-Correlation
Function (ICCF; Gaskell \& Sparke 1986 and Gaskell \& Peterson 1987 as
implemented by White \& Peterson 1994). In this approach one light curve is 
cross correlated with a linear interpolation of the second light curve, 
then the second light curve is cross correlated with a linear 
interpolation of the first light curve, and then the final cross 
correlation is the average of these two cross correlation functions.
The time lag is then determined as the centroid of all points in the ICCF with
correlation value above 80\% of the peak correlation value.

The second method is the $z$-transformed discrete correlation function 
(ZDCF: Alexander 1997) which is an improvement of the Discrete Correlation 
Function (DCF) method suggested by Edelson \& Krolik (1988). The ZDCF 
applies Fisher's $z$ transformation to the correlation coefficients, and 
uses equal population bins instead of the equal time bins that are used in 
the DCF. We also applied the JAVELIN method to our data. JAVELIN (Zu et al. 2013) 
uses the assumption that all emission-line light curves are scaled, smoothed, 
and displaced versions of the continuum. It fits the light curves directly 
using a damped random walk model and aligns them to recover the time lag 
and its statistical confidence limits. We find that with our detailed light 
curves JAVELIN produces similar results to the ICCF and ZDCF methods. 
In the following we refer only to the first two methods.

\begin{figure*}
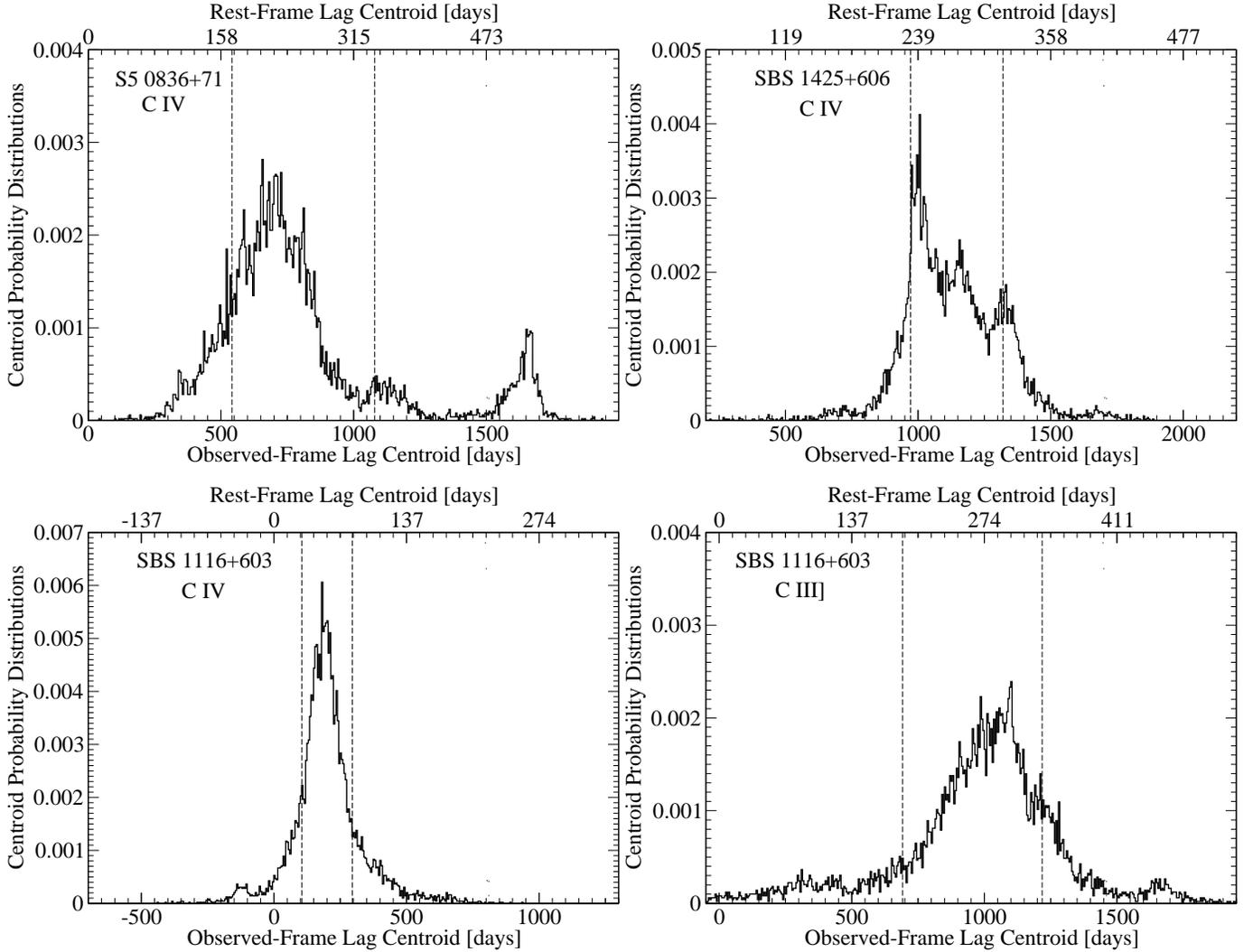

\gridline{\fig{fig8a.eps}{0.5\textwidth}{}
          \fig{fig8b.eps}{0.5\textwidth}{}
          }
\vspace{-0.7cm}
\gridline{\fig{fig8c.eps}{0.5\textwidth}{}
          \fig{fig8d.eps}{0.5\textwidth}{}
          }
\vspace{-0.5cm}
\caption{CCCDs for the four significant time lags found in this paper. The lag 
centroid is given in days in the observed frame and in the rest frame. The dashed vertical lines 
encompass the range which includes 68\% of the Monte-Carlo realizations. 
This range determines the uncertainty on the time-lag frond from the ICCF 
(Peterson et al. 1998).
\label{fig:CCCD}}
\end{figure*}

Figures~\ref{fig:s40636} to \ref{fig:hs1700} present the ICCF and ZDCF of 
the two lines (out of the three lines Ly$\alpha$, \ion{C}{3}] and 
\ion{C}{4}) which were monitored in each object. We consider a time lag as
significant\footnote{Further discussion of the significance is given at the 
end of sec.~\ref{subsec:summary}.} if the CCF possesses a maximum with peak 
correlation larger than 0.5 (e.g., Kaspi et al. 2000, Lira et al. 2018) in both 
the ICCF and ZDCF, and that it is a result of a significant overlap between the 
continuum and line light curves. We then calculated the uncertainties on the 
time lags of the ICCF method using the model-independent Monte Carlo method Flux
Randomization/Random Subset Selection (FR/RSS) of Peterson et al. (1998). 
This method yields a Cross-Correlation Centroid Distribution (CCCD) of the time lags
found in the Monte Carlo simulations of the light curves, and this distribution 
is used to estimate the uncertainties. The CCCDs for our time lags are
shown in Figure~\ref{fig:CCCD}.
The range of uncertainty is taken as the range which includes the central 
68\% of the CCCD realizations, i.e., the 16\% at the two edges of the
distribution are omitted and the central range determines the uncertainty
range.
This is demonstrated by the vertical dashed lines in Figure~\ref{fig:CCCD}.
Once the uncertainty range is determined the centroid of the ICCF, of the full 
two light curves, is taken as the time lag and its plus and minus uncertainties
are calculated to encapsulate this uncertainty range.

\ion{He}{2}\,$\lambda 1640$ is usually a weak line relative to 
\ion{C}{4}\,$\lambda 1550$ in high-luminosity AGNs. This is evident from the 
mean spectra shown in Fig.~\ref{fig:mean_std} where the line is absent or 
barely detected. This is also shown in Dietrich et al. (2002). Although they
find the Baldwin relation (Baldwin 1977) is about the same for both lines, the 
flux of the \ion{He}{2}\,$\lambda 1640$ line in their composite quasar spectra 
is much weaker than the \ion{C}{4}\,$\lambda 1550$ flux. 
The \ion{He}{2}\,$\lambda 1640$ line falls outside of the measured ranges 
listed in Table~\ref{tab:waves}, so this line does not affect our measured 
\ion{C}{4} time lag.

\subsection{Notes on Individual Objects}

The reported significant \ion{C}{4} time lags are listed in Table~\ref{civdata}.
Here we list notes on the time lags found in individual objects.

\subsubsection{S4\,0636+68}

No time lag is detected in the two emission lines that we monitored, Ly$\alpha$
and \ion{C}{4}.

\subsubsection{S5\,0836+71}

A \ion{C}{4} time lag of $188^{+27}_{-37}$ days in the quasar rest frame
was detected in this object as we previously 
reported in Paper~I. The current result is consistent with the result we
preliminarily found a decade ago when we had only
about half the data we are reporting in this work. 

The CCF of \ion{C}{3}] with the continuum in S5\,0836+71 shows a time lag 
of about $3570\pm 140$ days in the observed frame. Although formally 
significant, we note that this time lag is a result of the 
cross correlation of the first three years of the continuum light curve with 
the last three years of the \ion{C}{3}] light curve. This is too brief of an
overlap interval to detect significantly such a large time lag, and we do not
claim this detection as significant. This was also not detected in Paper~I due to 
the shorter monitoring period reported there.

Jorstad et al.(2013) reported an increase in the $\gamma$-ray activity of
S5\,0836+71 on March 2011 based on {\it Fermi} observations, 
which lasted for a year. They also report on a sharp $\gamma$-ray flare
at the end of 2011, which was also accompanied by an increase in the 
optical brightness of the quasar by $\sim 0.5$ mag in the R band. 
This is in accord with our light curve presented in Figure~\ref{fig:s50836} 
which shows that flare at the same time. No other such strong flares are seen in 
our light curves. Our result is not sensitive to this flare since it occurred about 
two years before the end of our campaign and any response to that continuum flare 
in the lines would be after the end of our observations.

\subsubsection{SBS\,1116+603}

A \ion{C}{4} time lag was detected for this object, though this was not detected 
in paper~I, probably due to the shorter monitoring period
and because there was no significant variability feature in the light curve 
in that period.

A \ion{C}{3}] time lag was also detected in this object and it is found to be 
$262^{+72}_{-72}$ days in the rest frame. This is the only
object where we detected a significant \ion{C}{3}] time lag out of the three 
objects in which we monitored that line.

\subsubsection{SBS\,1233+594}

There is a hint of a Ly$\alpha$ time lag in the ICCF at about 165 days in the
rest frame. However, this peak is not detected with the ZDCF method and using
the FR/RSS method the lag is consistent with zero. We therefore exclude this
result from further discussion. For the second line monitored in this object,
\ion{C}{4}, no time lag was detected.

\subsubsection{SBS\,1425+606}

A \ion{C}{4} time lag is detected but no Ly$\alpha$ time lag. In paper~I no
time lag was detected for this object as the continuum light curve in the first
10 years of the monitoring shows only a gradual rise, and a shorter-timescale
feature of luminosity increase and decrease was detected only in the last
eight years of monitoring.

\subsubsection{HS\,1700+6416}

Even though the continuum light curve shows variability and features
of higher and lower luminosity, no line variations are detected in response to
these continuum changes in the two emission lines that we monitored, Ly$\alpha$
and \ion{C}{4}. 

\subsubsection{Summary}
\label{subsec:summary}

We found line time lags for three out of the six monitored objects. Two of these
AGNs are radio loud and one is radio quiet (see Table~\ref{tab:luminosities}). 
In paper~I we reported on preliminary results from the first 5 years of monitoring, 
including one preliminary time lag measurement. No other lines showed 
statistically significant time lags in the first 5 years of monitoring. In this 
paper we find that after adding 8 more years to make a total of 13 years of
spectrophotometric monitoring we can measure time lags for three objects out of 
the six. This supports the idea that a monitoring period of more than a decade
is needed in order to carry out a successful reverberation-mapping campaign on
quasars at the high end of the luminosity range.

Half of the objects in our sample are radio loud and their continuum 
luminosity variability may be affected not only by the central accretion disk,
as is assumed in reverberation mapping, but also by the jet. Several 
studies (e.g., Le{\'o}n-Tavares et al. 2013; Paltani \& T{\"u}rler 2003) 
found evidence for components of the BLR that respond to jet flares seen 
in gamma-ray energies. However, the gamma-ray flares and the response seen
in emission lines like \ion{Mg}{2}, H$\alpha$ and H$\beta$, are on short 
time scales of days to weeks and our current study is not sensitive to such 
timescales due to our cadence of more than a month. Thus, the 
effect of possible jet flares on our time-lag results are negligible 
and probably well within our reported uncertainties.

We measured the rest-frame equivalent width of the lines in our spectra to be
10\AA\ to 20\AA\ for the \ion{C}{4} emission line and between 50\AA\ to 100 \AA\
for the Ly$\alpha$ emission line. These values are in accord with other 
high-redshift high-luminosity AGNs (e.g., Constantin et al. 2002; Dietrich et
al. 2002); this indicates that there is not large optical dilution by a
beamed jet optical continuum in these sources (e.g., Shaw et al. 2012).

Due to the small number of objects in our campaign and the small number of time 
lags which are detected, it is hard to find a way to estimate the number of
false detections in our study, which suffers from very small number statistics.
A basic way to estimate the number of false detections is to look at the
number of peaks on the negative side of the CCFs that could be declared as
significant time lags\footnote{As defined in sec.~\ref{sec:time}, we consider
a time lag as significant if the CCF possesses a maximum with peak 
correlation larger than 0.5 in both the ICCF and ZDCF, and that it is a result 
of a significant overlap between the continuum and line light curves.}. 
A careful inspection of our 12 CCFs yields that, if 
using the criteria to define a peak when it appears clearly both in the ICCF 
and ZDCF and has a peak coefficient above 0.5, then no feature on the negative
side of the CCF can be defined as a significant peak. This reinforces our 
detected positive lags as being true detections. 

\begin{deluxetable*}{lcccccc}
\tablecolumns{6}
\tabletypesize{\scriptsize}
\tablewidth{0pt}
\tablecaption{\ion{C}{4} data  \label{civdata}}
\tablehead{
\colhead{Object}        &
\colhead{$\lambda L_\lambda$(1350 \AA )} &
\colhead{Time lag\tablenotemark{\scriptsize a}}  &  
\colhead{reference} &
\colhead{redshift} &
\colhead{FWHM\tablenotemark{\scriptsize a}} &
\colhead{$\log{M_{BH}}$} \\
\colhead{}      &
\colhead{ergs\,s$^{-1}$} &
\colhead{days}  &
\colhead{}      &
\colhead{}      &
\colhead{\kms}  &
\colhead{$\log{\Msun}$} \\
\colhead{(1)} &
\colhead{(2)} &
\colhead{(3)} &
\colhead{(4)} &
\colhead{(5)} &
\colhead{(6)} &
\colhead{(7)} 
} 
\startdata
NGC\,4395 & $(8.15 \pm 0.51)\times 10^{39}$  & $0.040^{+0.024}_{-0.018}$ & Peterson et al. (2005) &0.001064&$3100\pm 1000$&$4.73^{+0.45}_{-0.61}$ \\
NGC\,3783 & $(3.89 \pm 0.81)\times 10^{43}$  & $3.8^{+1.0}_{-0.9}$       & Peterson et al. (2005) &0.00973&$3690\pm 480$& $6.88^{+0.21}_{-0.24}$ \\
NGC\,7469 & $(6.03 \pm 0.98)\times 10^{43}$  & $2.5^{+0.3}_{-0.2}$       & Peterson et al. (2005) &0.01632&$4310\pm 420$& $6.83^{+0.13}_{-0.13}$ \\
3C\,390.3 & $(1.18 \pm 0.59)\times 10^{44}$  & $35.7^{+11.4}_{-14.6}$    & Peterson et al. (2005) & 0.0561&$9700\pm 1700$&$8.69^{+0.26}_{-0.39}$ \\
NGC\,4151 & $(4.28 \pm 1.27)\times 10^{42}$  & $3.34^{+0.82}_{-0.77}$    & Metzroth et al. (2006) &0.00332&$5780\pm 920$& $7.21^{+0.22}_{-0.26}$ \\
NGC\,5548 & $(3.89 \pm 0.79)\times 10^{43}$  & $4.53^{+0.35}_{-0.34}$    & De Rosa et al.  (2015) &0.01676&$6710\pm 490$& $7.48^{+0.09}_{-0.10}$ \\
CTS\,286  & $(11.16\pm 1.83)\times 10^{46}$  & $459^{+71}_{-92}$         & Lira et al. (2018)     & 2.551 &$6260\pm 630$& $9.42^{+0.15}_{-0.19}$ \\
CTS\,406  & $(8.13 \pm 0.75)\times 10^{46}$  & $98^{+55}_{-74}$          & Lira et al. (2018)     & 3.178 &$6240\pm 620$& $8.75^{+0.28}_{-0.70}$ \\
CTS\,564  & $(9.95 \pm 1.51)\times 10^{46}$  & $115^{+184}_{-29}$        & Lira et al. (2018)     & 2.653 &$5620\pm 560$& $8.73^{+0.50}_{-0.22}$ \\
CTS\,650  & $(7.59 \pm 1.83)\times 10^{46}$  & $162^{+33}_{-10}$         & Lira et al. (2018)     & 2.659 &$3420\pm 340$& $8.44^{+0.16}_{-0.12}$ \\
CTS\,953  & $(9.99 \pm 1.97)\times 10^{46}$  & $73^{+115}_{-58}$         & Lira et al. (2018)     & 2.526 &$5000\pm 500$& $8.43^{+0.49}_{-0.78}$ \\
CTS\,1061 & $(33.88\pm 3.15)\times 10^{46}$  & $91^{+111}_{-24}$         & Lira et al. (2018)     & 3.368 &$3220\pm 320$& $8.14^{+0.43}_{-0.22}$ \\
J\,214355 & $(9.17 \pm 1.02)\times 10^{46}$  & $136^{+100}_{-90}$        & Lira et al. (2018)     & 2.607 &$6900\pm 690$& $8.98^{+0.32}_{-0.56}$ \\
J\,221516 & $(14.29\pm 1.86)\times 10^{46}$  & $153^{+91}_{-12}$         & Lira et al. (2018)     & 2.709 &$5890\pm 590$& $8.89^{+0.29}_{-0.13}$ \\
DES\,J0228-04 &$(2.69\pm 0.25)\times 10^{46}$ & $123^{+43}_{-42}$        &Hoormann et al. (2019)  & 1.905 &$7800\pm 1700$&$9.04^{+0.30}_{-0.40}$ \\
DES\,J0033-42 &$(3.24\pm 0.15)\times 10^{46}$ & $95^{+16}_{-23}$         &Hoormann et al. (2019)  & 2.593 &$7700\pm 650$& $8.92^{+0.14}_{-0.20}$ \\
RMID 032  & $(3.11 \pm 0.15)\times 10^{44}$   & $21.1^{+22.7}_{-8.3}$    &Grier et al. (2019)     & 1.715 &$5010\pm  20$& $7.89^{+0.32}_{-0.22}$ \\
RMID 052  & $(3.155\pm 0.015)\times 10^{45}$  & $32.6^{+6.9}_{-2.1}$     &Grier et al. (2019)     & 2.305 &$3354\pm  67$& $7.73^{+0.10}_{-0.05}$ \\
RMID 181  & $(3.508\pm 0.12)\times 10^{44}$   & $102.1^{+26.8}_{-10.0}$  &Grier et al. (2019)     & 1.675 &$4533\pm  49$& $8.49^{+0.11}_{-0.05}$ \\
RMID 249  & $(9.64 \pm 0.22)\times 10^{44}$   & $22.8^{+31.3}_{-11.5}$   &Grier et al. (2019)     & 1.717 &$2601\pm  29$& $7.35^{+0.38}_{-0.31}$ \\
RMID 256  & $(1.2270\pm 0.0085)\times 10^{45}$& $43.1^{+49.0}_{-15.1}$   &Grier et al. (2019)     & 2.244 &$3565\pm   9$& $7.90^{+0.34}_{-0.20}$ \\
RMID 275  & $(4.0830\pm 0.0094)\times 10^{45}$& $76.7^{+10.0}_{-3.9}$    &Grier et al. (2019)     & 1.577 &$6943\pm  22$& $8.73^{+0.06}_{-0.03}$ \\
RMID 298  & $(3.9450\pm 0.0091)\times 10^{45}$& $82.3^{+64.5}_{-24.5}$   &Grier et al. (2019)     & 1.635 &$5177\pm  51$& $8.51^{+0.26}_{-0.16}$ \\
RMID 312  & $(1.194\pm 0.011)\times 10^{45}$  & $70.9^{+9.6}_{-3.3}$     &Grier et al. (2019)     & 1.924 &$10248\pm 53$& $9.04^{+0.06}_{-0.03}$ \\
RMID 332  & $(3.556\pm 0.016)\times 10^{45}$  & $83.8^{+23.3}_{-6.5}$    &Grier et al. (2019)     & 2.581 &$7828\pm  32$& $8.88^{+0.11}_{-0.04}$ \\
RMID 387  & $(4.864\pm 0.011)\times 10^{45}$  & $48.4^{+34.7}_{-10.1}$   &Grier et al. (2019)     & 2.426 &$4797\pm  30$& $8.21^{+0.24}_{-0.11}$ \\
RMID 401  & $(3.090\pm 0.021)\times 10^{45}$  & $60.6^{+36.7}_{-13.0}$   &Grier et al. (2019)     & 1.822 &$10120\pm 497$&$8.96^{+0.25}_{-0.15}$ \\
RMID 418  & $(1.0960\pm 0.0076)\times 10^{45}$& $58.6^{+51.6}_{-21.3}$   &Grier et al. (2019)     & 1.418 &$6159\pm  44$& $8.51^{+0.28}_{-0.20}$ \\
RMID 470  & $(6.622\pm 0.092)\times 10^{44}$  & $27.4^{+63.5}_{-22.0}$   &Grier et al. (2019)     & 1.879 &$5028\pm  70$& $8.01^{+0.53}_{-0.72}$ \\
RMID 527  & $(6.138\pm 0.042)\times 10^{44}$  & $47.3^{+13.3}_{-5.0}$    &Grier et al. (2019)     & 1.647 &$8306\pm  53$& $8.68^{+0.11}_{-0.05}$ \\
RMID 549  & $(2.339\pm 0.011)\times 10^{45}$  & $68.9^{+31.6}_{-9.6}$    &Grier et al. (2019)     & 2.275 &$4995\pm  53$& $8.40^{+0.17}_{-0.07}$ \\
RMID 734  & $(3.3880\pm 0.0078)\times 10^{45}$& $68.0^{+38.2}_{-11.5}$   &Grier et al. (2019)     & 2.332 &$7042\pm  65$& $8.69^{+0.20}_{-0.09}$ \\
RMID 363  & $3.16\pm 0.15)\times 10^{46}$     & $300.4^{+17.1}_{-4.7}$   &Shen et al. (2019)      & 2.635 &$ 5252\pm 94$\tablenotemark{b}& $9.08^{+0.04}_{-0.02}$ \\       
RMID 372  & $4.17\pm 0.19)\times 10^{45}$     & $67.0^{+20.4}_{-7.4}$    &Shen et al. (2019)      & 1.745 &$10451\pm 68$\tablenotemark{b}& $9.03^{+0.12}_{-0.06}$ \\       
RMID 651  & $2.63\pm 0.12)\times 10^{45}$     & $91.7^{+56.3}_{-22.7}$   &Shen et al. (2019)      & 1.486 &$ 6391\pm 52$\tablenotemark{b}& $8.74^{+0.21}_{-0.13}$ \\      
S5\,0836+71  & $(1.03\pm 0.16)\times 10^{47}$ & $230^{+91}_{-59}$       & This work              & 2.172 &$9000\pm 800$ &$9.44^{+0.22}_{-0.21}$ \\ 
SBS\,1116+603& $(1.80\pm 0.20)\times 10^{47}$ & $65^{+17}_{-37}$        & This work              & 2.646 &$4800\pm 450$ &$8.34^{+0.18}_{-0.45}$ \\
SBS\,1425+606& $(4.52\pm 0.37)\times 10^{47}$ & $285^{+30}_{-53}$       & This work              & 3.192 &$8200\pm 470$ &$9.45^{+0.09}_{-0.14}$ 
\enddata
\tablenotetext{a}{In the rest frame.}
\tablenotetext{a}{Computed from the measured $\sigma_{line}$ using the line profile Gaussian assumption: FWHM$=2.355\times\sigma_{line}$.} 
\end{deluxetable*}

\begin{figure*}
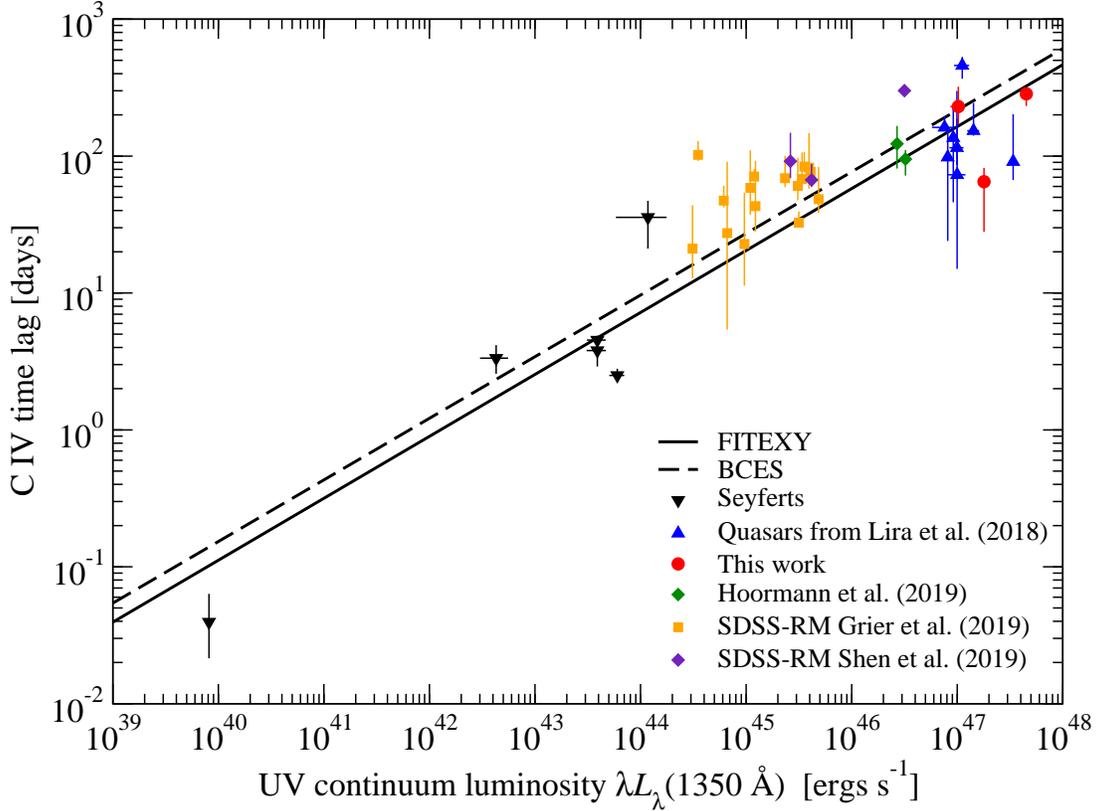

\gridline{\fig{fig9.eps}{0.8\textwidth}{}
          }
\caption{\ion{C}{4} BLR size --- UV Luminosity relation. 
Time lag is in the quasar rest frame.\label{RvsLuv}}
\end{figure*}

\section{Disscussion} 
\label{sec:diss}

\subsection{\ion{C}{4} BLR Size --- Luminosity Relation}

In Table~\ref{civdata}, we list the three objects for which we have found 
significant and reliable \ion{C}{4} reverberation lags. To these we add a 
compilation of previous measurements, including \ion{C}{4} BLR sizes for 
a few Seyfert AGNs which were reanalyzed and compiled by Peterson
et al. (2004, 2005), Metzroth et al. (2006), and De Rosa et al. (2015).
Lira et al (2018) reported time lags for eight high-luminosity quasars,
but with a number of mistakes which were corrected in a recent erratum (e.g., 
a few wrong rest-frame time lags and their uncertainties in their 
table~4 and mismatched objects in their figure~7); these mistakes were only
in the presentation and did not affect their final analysis. The numbers
in Table~\ref{civdata} include all the corrections to the Lira et al (2018)
measurements. We also include in Table~\ref{civdata} the two quasars reported by 
Hoormann et al. (2019), and also the 16 objects which are the ``gold sample''
of Grier et al. (2019), and the three objects from Shen et al. (2019).
Figure~\ref{RvsLuv} displays the 38 objects from Table~\ref{civdata} with
reliable and significant \ion{C}{4} time-lag measurements.

Linear regression for the \ion{C}{4} BLR size --- Luminosity Relation is performed
with the two methods detailed in Kaspi et al. (2005, 2007): (1) the FITEXY approach
of Press et al. (1992, p. 660) with the Tremaine et al. (2002) procedure to 
account for the intrinsic scatter in the data, and (2) the bivariate correlated 
errors and intrinsic scatter (BCES) regression method (Akritas \& Bershady 1996).
The two techniques take into account the uncertainty in the BLR size, the 
luminosity, and the intrinsic scatter around a straight line.
The fit from the FITEXY method (solid line in Fig.~\ref{RvsLuv}) is
\begin{equation} \label{eq:RLa}
\frac{R_{BLR}}{10 {\rm \ lt\,days}}= (0.25\pm0.05)\left(\frac{\lambda
L_{\lambda}(1350\,{\mbox{\rm \AA}} )}{\rm 10^{43} \ ergs\,s^{-1}}
\right)^{0.45\pm0.03} ,
\end{equation}
and for the BCES method (dashed line in Fig.~\ref{RvsLuv}) it is
\begin{equation}
\frac{R_{BLR}}{10 {\rm \ lt\,days}}= (0.34\pm0.11)\left(\frac{\lambda L_{\lambda}(1350\,{\mbox{\rm \AA}} )}{\rm 10^{43} \ ergs\,s^{-1}} \right)^{0.45\pm0.05} .
\end{equation}
For these data the Pearson correlation coefficient is 0.90 with significance level
of $4\times 10^{-13}$ and the Spearman rank-order correlation is 0.80
with a significance level of $1\times 10^{-10}$. The intrinsic scatter using
the Tremaine et al. (2002) procedure is 58\%. These results are similar to
those found by Lira et al. (2018; a slope of $0.46\pm 0.08$
and a constant of $0.22\pm 0.10$) and
are in accord with the preliminary results reported in Kaspi et al. (2007).
The uncertainties are about half of these reported by Lira et al. (2018).
Our slope is also consistent, within the uncertainties, with that measured by 
Grier et al. (2019) who find a slope of $0.51\pm 0.05$ for their full sample,
and a slope of $0.52\pm 0.04$ for their gold sample.

Almost all objects in Figure~\ref{RvsLuv} lie in the range of 
$ {\rm 10^{43}}\lesssim \lambda L_{\lambda}(1350\,{\mbox{\rm \AA}})\lesssim {\rm 10^{48} \ ergs\,s^{-1}}$, i.e., covering five orders of magnitudes in luminosity. 
There is one object three orders
of magnitudes below that range: NGC\,4395 at 
$\lambda L_{\lambda}(1350\,{\mbox{\rm \AA}}) \approx {\rm 10^{40} \ ergs\,s^{-1}}$. Thus, this one object may via its statistical leverage influence significantly the 
BLR size --- Luminosity relation. Repeating the fit without NGC\,4395 in the
sample we find a slope of $0.40\pm 0.04$ using the FITEXY method and 
$0.38\pm 0.05$ using the BCES method. This may indicate that the true relation
has a shallower slope than the theoretically predicted slope of 0.5,
though further reverberation studies
on lower-luminosity AGNs are needed in order to better understand this issue.  

We have found a significant \ion{C}{4} time lag for three out of our six 
spectroscopically monitored quasars, i.e., in 50\% of the objects studied. 
This success rate is comparable to that of Lira et al. (2018) who found
\ion{C}{4} time lags for eight out of their 17 monitored quasars. 
In contrast, while Lira et al. (2018) found also
Ly$\alpha$ time lags in 50\% of their objects, we have found only a hint of a
Ly$\alpha$ time lag in one of our objects. This is probably because in two of
our objects only part of the Ly$\alpha$ line is within the spectral 
range covered, in one it is outside the range, and 
in the other two cases the light curve is too noisy due to low S/N at the 
blue end of the spectrum and the many absorption lines from the 
Ly$\alpha$ forest. We measure one significant \ion{C}{3}] time lag, as did
Lira et al. (2018). This line is much weaker than \ion{C}{4} and hence its 
light curves are noisy and challenging for time-lag detection.

All six AGNs in our spectrophotometrically monitored sample show 
significant 1350\AA\ continuum variations at the level of 20\%--70\%. However, 
we detect corresponding \ion{C}{4} variations only in half of the them. 
This is similar to Lira et al. (2018) who also do not find corresponding
variations in the \ion{C}{4} light curves of about half of their AGNs sample,
in spite of significant variations in their continuum light curves. 
This may suggest that the ionizing continuum in high-luminosity AGNs behaves
differently from the 1350\AA\ continuum and that this different behavior is
a common phenomenon in these objects. 

Comparing the H$\beta$ BLR size to the \ion{C}{4} BLR size is important for
constraining BLR models which aim to produce the BLR emission lines (e.g., 
Netzer 2020). The calculated BLR models predict that the \ion{C}{4} BLR size
is about half the H$\beta$ BLR size.
Using the measured relation of H$\beta$ BLR size and the luminosity 
$\lambda L_\lambda$(5100\AA ) of the entire sample in eq.~5 of Du et al. 
(2015) and the estimated $\lambda L_\lambda$(5100\AA ) of the objects in 
our sample (Table~\ref{tab:luminosities})\footnote{Note that these 
$\lambda L_\lambda$(5100\AA ) estimates are based on our measured 
$\lambda L_\lambda$(1350\AA ) and thus they are not independent
measurements.}, we can estimate the H$\beta$ BLR 
size for the three objects in which we measured the \ion{C}{4} BLR size.
We find H$\beta$ BLR sizes of 
$650^{+210}_{-160}$ light days for S5\,08036+71,
$855^{+295}_{-220}$ light days for SBS\,1116+603, and
$1360^{+520}_{-380}$ light days for SBS\,1425+606.
The ratio between these sizes and the \ion{C}{4} BLR sizes measured 
in this work shows a large scatter between 3 to 13. This is mainly due to the 
exceptionally small \ion{C}{4} BLR size which we measured in SBS\,1116+603.
Also the fact that we use only three objects and not a larger sample does not
enable proper statistics with small standard deviation.
Thus, in order to compare the H$\beta$ and \ion{C}{4} BLR sizes, it is better
to use the mean relations. We therefore compare the mean relation between the 
\ion{C}{4} BLR size and $\lambda L_\lambda$(1350\AA ) which we find 
in eq.~\ref{eq:RLa} to the mean relation between the H$\beta$ BLR size
and $\lambda L_\lambda$(5100\AA ) from Du et al. (2015).
Using these averaged measured BLR sizes we find that
for $\lambda L_\lambda$(5100\AA )=$10^{45}$\,erg\,s$^{-1}$ the ratio between 
the H$\beta$ to \ion{C}{4} BLR sizes is 3.3, while for 
$\lambda L_\lambda$(5100\AA )=$10^{47}$\,erg\,s$^{-1}$ this factor is 4.1 .
This is larger than previously predicted by BLR models and thus a change in the 
models (Netzer 2020) may be needed in order to fit the observations.

\subsection{Mass --- Luminosity Relation}

Assuming gravitationally dominated motions of the BLR clouds, the central masses
of AGNs can be estimated using $M_{BH} = fG^{-1}V^2r$ where $V$ is an estimate
of the velocity of the BLR around the central mass and $r$ is an
estimate of the typical distance between the BLR gas and the central BH. 
$f$ is a scaling factor which embodies our ignorance of the BLR geometry 
and velocity field (e.g., Peterson et al. 2004). 
This method to estimate $M_{BH}$ has been widely used, over the past two 
decades, with various choices for $f$ and for the observational 
indicator of $v$, mainly for BLR distances determined from reverberation
mapping of the H$\beta$ emission line.

Although several attempts have been made to estimate $M_{BH}$ from
\ion{C}{4} emission-line reverberation (e.g., Vestergaard \& Peterson 2006; 
Park et al. 2013) it was found that this line has several drawbacks. The width of 
the \ion{C}{4} line is weakly correlated with the width of low-ionization lines
like H$\beta$ and \ion{Mg}{2}. Also, the \ion{C}{4} profiles often show large 
blueshifts and asymmetries that indicate non-virial motions (e.g., Baskin \& 
Laor 2005; Netzer et al. 2007; Shang et al. 2007; Richards et al. 2011; 
Trakhtenbrot \& Netzer 2012). 
Coatman et al. (2017) used a sample of 230 quasars to quantify 
the bias in \ion{C}{4} black-hole masses as a function of the \ion{C}{4} blueshift
and suggest an empirical formula to derive the black-hole mass from the 
\ion{C}{4} emission line properties.
On the other hand, Mej{\'\i}a-Restrepo et al. (2018) further studied the
possibility that \ion{C}{4} can serve for black-hole mass estimates from a
single-epoch spectrum by using three different methods to improve the 
measurement. They find these methods to be of limited applicability, mostly 
because they depend on correlations that are driven by the line width of
the \ion{C}{4} profile and not by an interconnection between the line width
of the \ion{C}{4} line and the line width of the low-ionization lines. 
The conclusion of Mej{\'\i}a-Restrepo et al. (2018) is that \ion{C}{4}-based
mass estimates at high redshift cannot serve as an alternative for estimates
based low-ionization lines like H$\alpha$, H$\beta$, and \ion{Mg}{2}.

Mej{\'\i}a-Restrepo et al. (2016) used a sample of 39 AGNs at $z\sim 1.55$ 
that have measurements of both the H$\beta$ and \ion{C}{4} emission lines to
measure black-hole mass from the two lines. In addition, Vietri et al. (2020)
used a sample of 21 AGN at $z\sim 2$ with measurements of these two lines
to derive the black-hole masses from the two lines. In both studies the agreement
between the derived masses from the two lines is poor, and this confirms the
inconsistency of mass estimates based on the \ion{C}{4} emission line.

Dalla Bont\'a et al. (2020) study black-hole masses in AGNs based on single-epoch
spectra. They used the compilation of reverberation-mapped AGNs by Bentz \& 
Katz (2015) and the AGNs from the SDSS-RM project as well. These authors find 
that the line dispersion of the \ion{C}{4} emission line is a better proxy for 
estimating the black-hole mass than the FWHM. They find that, in addition to
luminosity and line width, a third parameter is required to obtain accurate
masses and that parameter seems to be the Eddington ratio, and they present
empirical relations for estimating black-hole masses from the H$\beta$ and
\ion{C}{4} emission lines.

Nevertheless, since for high-luminosity, high-redshift ($z\sim 3$), quasars
there are no low-ionization line reverberation-mapping measurements, and most
measurements which can be obtained are of the \ion{C}{4} line, it is
interesting to estimate the black-hole mass and to check its correlation with
the luminosity, even given the above limitations. We note that from our mean
spectra (Figure~\ref{fig:mean_std}) we find that the \ion{C}{4} emission line
is not broader and not blueshifted relative to the Ly$\alpha$ emission line.
Thus, the mass estimates of the objects in our sample may be more reliable as 
our objects are free from some of the above limitations which discourage the 
use of \ion{C}{4} for mass estimates.

\begin{figure*}
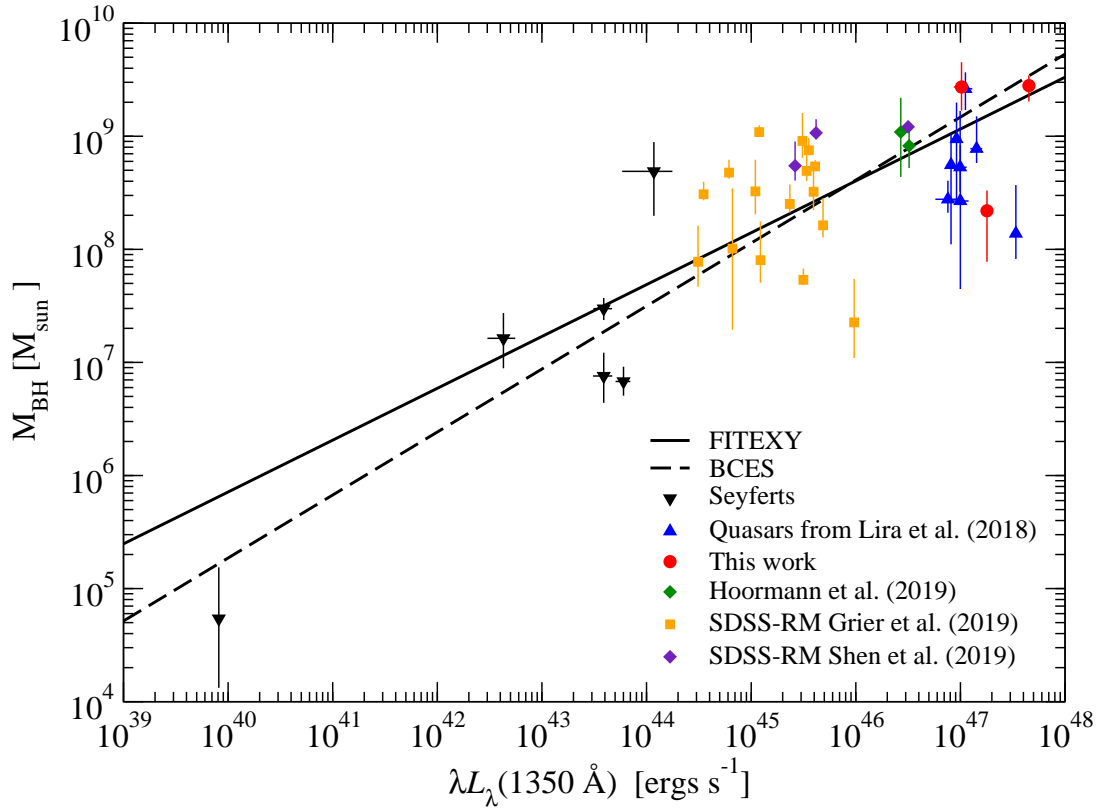

\gridline{\fig{fig10.eps}{0.8\textwidth}{}
          }
\caption{Black hole mass estimate --- UV Luminosity relation. \label{mass}}
\end{figure*}

For simplicity and given the very basic assumptions, we use the equation given
by Kaspi et al. (2000; Equation~5 therein) where the FWHM of the line is used
as the BLR clouds' velocity and it is corrected by a factor of $\sqrt{3}/2$,
to account for velocities in three dimensions. The FWHMs we use are listed in
Table~\ref{civdata} together with the derived masses. Fig.~\ref{mass} shows
the black-hole mass and UV luminosity relations for the 38 AGNs listed in
Table~\ref{civdata}. 
A fit using the FITEXY method (solid line in Fig.~\ref{mass}) is

\begin{equation}
\frac{M_{BH}}{10^7 {\rm \ M_{\odot}}}= (1.70^{+0.99}_{-0.58})\left(\frac{\lambda L_{\lambda}(1350\,{\mbox{\rm \AA}} )}{\rm 10^{43} \ ergs\,s^{-1}} \right)^{0.45\pm0.06} ,
\end{equation}
and for the BCES method (dashed line in Fig.~\ref{mass}) it is
\begin{equation}
\frac{M_{BH}}{10^7 {\rm \ M_{\odot}}}= (0.87^{+0.35}_{-0.25})\left(\frac{\lambda L_{\lambda}(1350\,{\mbox{\rm \AA}} )}{\rm 10^{43} \ ergs\,s^{-1}} \right)^{0.56\pm0.05} .
\end{equation}
For these data the Pearson correlation coefficient is 0.79 with significance
level of $5.4\times 10^{-9}$ and the Spearman rank-order correlation is 0.56
with a significance level of $2.5\times 10^{-4}$. The intrinsic scatter using
the Tremaine et al. (2002) procedure is 240\%. 

Theoretically we expect $M_{BH} \propto L^{0.5}$ which 
emerges since $M_{BH} \propto V^2r$
and because it is found that $r \propto L^{0.5}$, thus also the black-hole mass
will be proportional to $L^{0.5}$. Since the velocity of the BLR (taken here
as the FWHM of the line) does not evolve with the luminosity it only introduces
scatter in the black-hole mass for a given luminosity; thus the scatter in
this relation is larger than the scatter found earlier between the size of the
BLR and the luminosity. It should also be noted that the sample in 
Table~\ref{civdata} does not represent the AGN population in luminosity or in 
black-hole mass; thus, the relations found in equations 3 and 4 
may relate more to way the sample 
was selected, i.e., all objects with significant \ion{C}{4} BLR size measurements.

For five Seyfert AGNs out of the six in Table~\ref{civdata} there are 
black-hole mass estimates from H$\beta$ reverberation mapping.
In Table~\ref{tab:Hb_MBH} we list these black-hole mass estimates 
as given in the database of 
Bentz \& Katz (2015)\footnote{http://www.astro.gsu.edu/AGNmass/}. 
From a comparison of these mass estimates
to our estimates from the \ion{C}{4} line in Table~\ref{civdata} we find that
for NGC\,7469 and 3C\,390.3 the black-hole mass estimates from the
two lines are within the 1$\sigma$ uncertainties (i.e., the 1$\sigma$ 
uncertainties overlap). For NGC\,3783 and NGC\,4151 the black-hole
mass estimates are within the 1.5$\sigma$ uncertainties (if we add the $\sigma$
from both measurements), and for NGC\,5548 they
are within 2$\sigma$. This resemblance in the mass estimates from the two
emission lines increases the confidence in these results and in the ability
of the reverberation-mapping method to give consistent mass estimates.

\begin{deluxetable}{cc}
\tablecolumns{2}
\tablewidth{0pt}
\tablecaption{Black-hole mass estimates from H$\beta$ reverberation mapping
\label{tab:Hb_MBH}}
\tablehead{
\colhead{Object}    &
\colhead{$\log{M_{BH}}$}
} 
\startdata
  NGC\,3783 & \ \ \ \ \ \ \ \ \ \ \ \ \ \ \ \ \ \ $7.42^{+0.13}_{-0.14}$ \\
  NGC\,7469 & \ \ \ \ \ \ \ \ \ \ \ \ \ \ \ \ \ \ $6.99^{+0.09}_{-0.01}$ \\
  3C\,390.3 & \ \ \ \ \ \ \ \ \ \ \ \ \ \ \ \ \ \ $8.48^{+0.06}_{-0.08}$ \\
  NGC\,4151 & \ \ \ \ \ \ \ \ \ \ \ \ \ \ \ \ \ \ $7.60^{+0.07}_{-0.06}$ \\
  NGC\,5548 & \ \ \ \ \ \ \ \ \ \ \ \ \ \ \ \ \ \ $7.72^{+0.02}_{-0.02}$ \\
\enddata
\footnotesize
\tablecomments{Measurements from the database of Bentz \& Katz (2015) in units
of $\log{\Msun}$. Note that the
front page of the database lists the averaged mass measurements from all lines
and here we list the mass measurements from the H$\beta$ line only.}
\end{deluxetable}

\section{Summary} \label{sec:sum}

We have reported the final results from a 13-year spectrophotometric 
reverberation-mapping campaign on high-luminosity quasars. 
The spectrophotometric observations were supplemented by 18 years of 
photometric monitoring, which improved the cadence and the accuracy of the 
spectrophotometric continuum light curves, by adding to them the photometric light curves.

All 11 objects with photometric light curves display continuum variability 
of 10\%---70\% over about six years in the rest frame. 
This value is similar to that reported in paper~I with data over a period which 
was about a third of the current one. Six of the 11 objects were 
monitored spectroscopically. We identified \ion{C}{4} emission-line time lags
for three objects indicating \ion{C}{4} emission-line region sizes of order 
100 to 250  light days. Together with data from previous studies, we have
constructed a \ion{C}{4} BLR size --- UV luminosity relation over
eight orders of magnitude in UV luminosity; the slope of $0.45\pm0.04$ is
consistent with previous studies and with photoionization theory.

Although all our monitored AGNs show significant continuum variation, only in 
about half of them we measured \ion{C}{4} time lag (same fraction was also was
found by Lira et al. 2018). This may indicate that the ionizing continuum 
behaves differently from the 1350\AA\ continuum we measured in this study.

Although \ion{C}{4} is not a preferable line for the central black-hole
mass determination, we derive the mass for the 19 objects with proper \ion{C}{4}
reverberation mapping data. We find that the mass scales approximately
as the square root of the UV luminosity, but with a large scatter. The
scatter probably results, at least partly, from the drawbacks of 
the \ion{C}{4} line for black-hole mass measurement, but probably also includes an 
intrinsic scatter in that relation for AGNs, e.g., different velocities for given
BLR size.

Our results demonstrate that for reverberation mapping of high-luminosity quasars
to succeed a long monitoring period is needed, of order a decade or more, 
and a large enough sample is required in order to be able to detect the 
responses to continuum variations in at least some of the objects in the sample.
Curent studies for \ion{C}{4} reverberation mapping cover mostly the UV 
luminosity range above $10^{43}$\,erg\,s$^{-1}$ and more studies are needed
to cover the lower-luminosity range of the AGNs phenomenon.

\acknowledgments

\vglue0.4cm
We are grateful to the staffs of WO and HET for their great assistance in
executing this long-term program. Special thanks go to John Dann,
Ezra Mashal, and Sami Ben-Gigi of the WO and to Gary Hill of the HET
for devoted technical support of this project through the years.
The Hobby-Eberly Telescope (HET) is a joint project of the
University of Texas at Austin, the Pennsylvania State University,
Stanford University, Ludwig-Maximillians-Universit\"at M\"unchen,
and Georg-August-Universit\"at G\"ottingen. The HET is named in
honor of its principal benefactors, William P. Hobby and Robert
E. Eberly. The Marcario Low-Resolution Spectrograph is named for
Mike Marcario of High Lonesome Optics, who fabricated several optics
for the instrument but died before its completion; it is a joint
project of the Hobby-Eberly Telescope partnership and the Instituto
de Astronom\'{\i}a de la Universidad Nacional Aut\'onoma de M\'exico.
We gratefully acknowledge the financial support of 
the Colton Foundation at Tel-Aviv University (S.~K.), NASA ADP grant 
80NSSC18K0878 and the V.M. Willaman Endowment (W.~N.~B).
This research has made use of the NASA/IPAC Extragalactic Database
(NED) which is operated by the Jet Propulsion Laboratory, California
Institute of Technology, under contract with the National Aeronautics
and Space Administration.

\appendix

\section{Photometric light curves and analysis}
\label{appendix}

Figures~\ref{fig:phot_lc_R2} and \ref{fig:phot_lc_B2} and 
Table~\ref{tab:phot_lcRB} present the $R$ and $B$ band light curves of the 
five quasars which were observed only photometrically at the Wise Observatory. 
Also, Figure~\ref{fig:phot_lc_B1} and Table~\ref{tab:phot_lcB} 
present the $B$ band light curves for the 
six spectroscopically monitored quasars for which their $R$ band light curves were 
merged into the spectroscopic continuum light curves and are shown in 
Figures~\ref{fig:s40636} to~\ref{fig:hs1700}

In several cases multiple observations of a given quasar were obtained 
during the same night; these measurements were averaged 
into one point in the light curve. 
Apparent magnitude calibration was achieved by using non-variable stars in the
field of each quasar with $B$ and $R$ magnitudes taken from the USNO-A2.0 
catalog.\footnote{see: http://brucegary.net/dummies/USNO-A2\_Method.htm}
The number of observations as well as statistical information about the light
curves are provided at the bottom of Table~\ref{vartab}.

We have cross correlated the photometric $B$ and $R$ band light curves for all
objects. The CCF results are listed in Table~\ref{Tab:BR} and two examples of
CCFs are shown in Figure~\ref{fig:BRccf}. The significance of the results is
not high since the monthly sampling rate limits the detected time lags. Also,
there is probably a mixture of lines and continuum in some of these broad bands.
Such mixture a will cause the detected variations in the broad bands to be
superpositions of the continuum variations and the line variations, which are
not in phase. This limits the ability to determine the time lags between the
red continuum and the blue continuum in these AGNs. Nevertheless, three objects
(SBS\,1116+603, SBS\,1425+606, and S5\,2017+744) show, formally, negative time
lags both in their centroid and the peak of the CCF. This formally means that
the $R$ band light curve is lagging the $B$ band light curve. Although we do
not refer to these results as highly significant, this is in accordance with
results from other reverberation-mapping studies which found that the red
continuum emission of AGNs is lagging the blue continuum emission (e.g.,
Collier et al. 1998, Edelson et al. 2019, Chelouche et al. 2019).

\begin{deluxetable}{cc}
\tablecolumns{2}
\tablewidth{0pt}
\tablecaption{$R$ and $B$ band light curves for five objects
\label{tab:phot_lcRB}}
\tablehead{
\colhead{JD}    &
\colhead{Apparent Magnitude}
} 
\startdata
\multicolumn{2}{c}{S5\,0014+81 $R$ band}   \\
\tableline
  2449667.28 & $ 16.843 \pm 0.031 $ \\
  2449728.24 & $ 16.860 \pm 0.016 $ \\
  2449918.57 & $ 16.850 \pm 0.019 $ \\
  2449929.52 & $ 16.864 \pm 0.016 $ \\
  2449954.46 & $ 16.864 \pm 0.031 $ \\
  2449981.42 & $ 16.852 \pm 0.023 $ \\
  2450007.39 & $ 16.888 \pm 0.020 $ \\
  2450034.32 & $ 16.857 \pm 0.026 $ \\
  2450076.22 & $ 16.842 \pm 0.023 $ \\
  2450104.20 & $ 16.839 \pm 0.019 $ \\
\enddata
\footnotesize
\tablecomments{Table~\ref{tab:phot_lcRB} is presented in its
entirety in the electronic edition of the Astrophysical Journal. A
portion is shown here for guidance regarding its form and content. }
\end{deluxetable}

\begin{deluxetable}{cc}
\tablecolumns{2}
\tablewidth{0pt}
\tablecaption{$B$ band light curves for six objects
\label{tab:phot_lcB}}
\tablehead{
\colhead{JD}    &
\colhead{Apparent Magnitude}
} 
\startdata
\multicolumn{2}{c}{S5\,0636+68 $B$ band}   \\
\tableline
     2449667.43   & $      17.160   \pm      0.024 $ \\
     2449728.43   & $      17.119   \pm      0.032 $ \\
     2449773.25   & $      17.134   \pm      0.028 $ \\
     2449804.22   & $      17.136   \pm      0.027 $ \\
     2449954.57   & $      17.216   \pm      0.032 $ \\
     2449981.59   & $      17.191   \pm      0.025 $ \\
     2450007.51   & $      17.241   \pm      0.028 $ \\
     2450035.48   & $      17.232   \pm      0.029 $ \\
     2450080.23   & $      17.267   \pm      0.027 $ \\
\enddata
\footnotesize
\tablecomments{Table~\ref{tab:phot_lcB} is presented in its
entirety in the electronic edition of the Astrophysical Journal. A
portion is shown here for guidance regarding its form and content. }
\end{deluxetable}

\begin{deluxetable}{lcc}
\tablecaption{Time lags of ICCF between $B$ and $R$ bands  \label{Tab:BR}}
\tablecolumns{3}
\tablewidth{0pt}
\tablehead{
\colhead{Object}    &
\colhead{$\tau_{\rm Centroid}$} &
\colhead{$\tau_{\rm Peak}$}  }
\startdata
S5 0014+81   & $ -4^{+59}_{-76}$  & $ 6^{+59}_{-106}$  \\
S5 0153+74   & $ 46^{+199}_{-71}$ & $ -18^{+118}_{-32}$  \\
S4 0636+68   & $ 6^{+49}_{-51}$ & $ 4^{+66}_{-39}$  \\
S5 0836+71   & $ 63^{+27}_{-43}$ & $ 6^{+29}_{-11}$  \\
TB 0933+733  & $ 2^{+43}_{-67}$ & $ 2^{+48}_{-62}$  \\
SBS 1116+603 & $ -145^{+85}_{-65}$ & $ -114^{+89}_{-26}$  \\
SBS 1233+594 & $ -43^{+63}_{-72}$ & $ -16^{+36}_{-74}$  \\
SBS 1425+606 & $ -264^{+39}_{-146}$ & $ -154^{+134}_{-181}$  \\
HS 1700+6416 & $ -75^{+55}_{-20}$ & $ 6^{+4}_{-81}$  \\
HS 1946+7658 & $ -33^{+78}_{-72}$ & $ 8^{+37}_{-118}$  \\
S5 2017+744  & $ -87^{+62}_{-58}$ & $ -108^{+103}_{-52}$  \\
\enddata
\tablecomments{Time lags are given in days in the observed frame.
A minus sign before the time lag means that formally the $R$ band is lagging 
behind the $B$ band.}
\end{deluxetable}

\begin{figure*}
\gridline{\fig{fig11.eps}{1.0\textwidth}{}
          }
\caption{$R$-band light curves of the five quasars that have no spectroscopic observations.
\label{fig:phot_lc_R2}
}
\end{figure*}

\begin{figure*}
\gridline{\fig{fig12.eps}{1.0\textwidth}{}
          }
\caption{$B$-band light curves of the five quasars that have no 
spectroscopic observations.
\label{fig:phot_lc_B2}}
\end{figure*}

\begin{figure*}
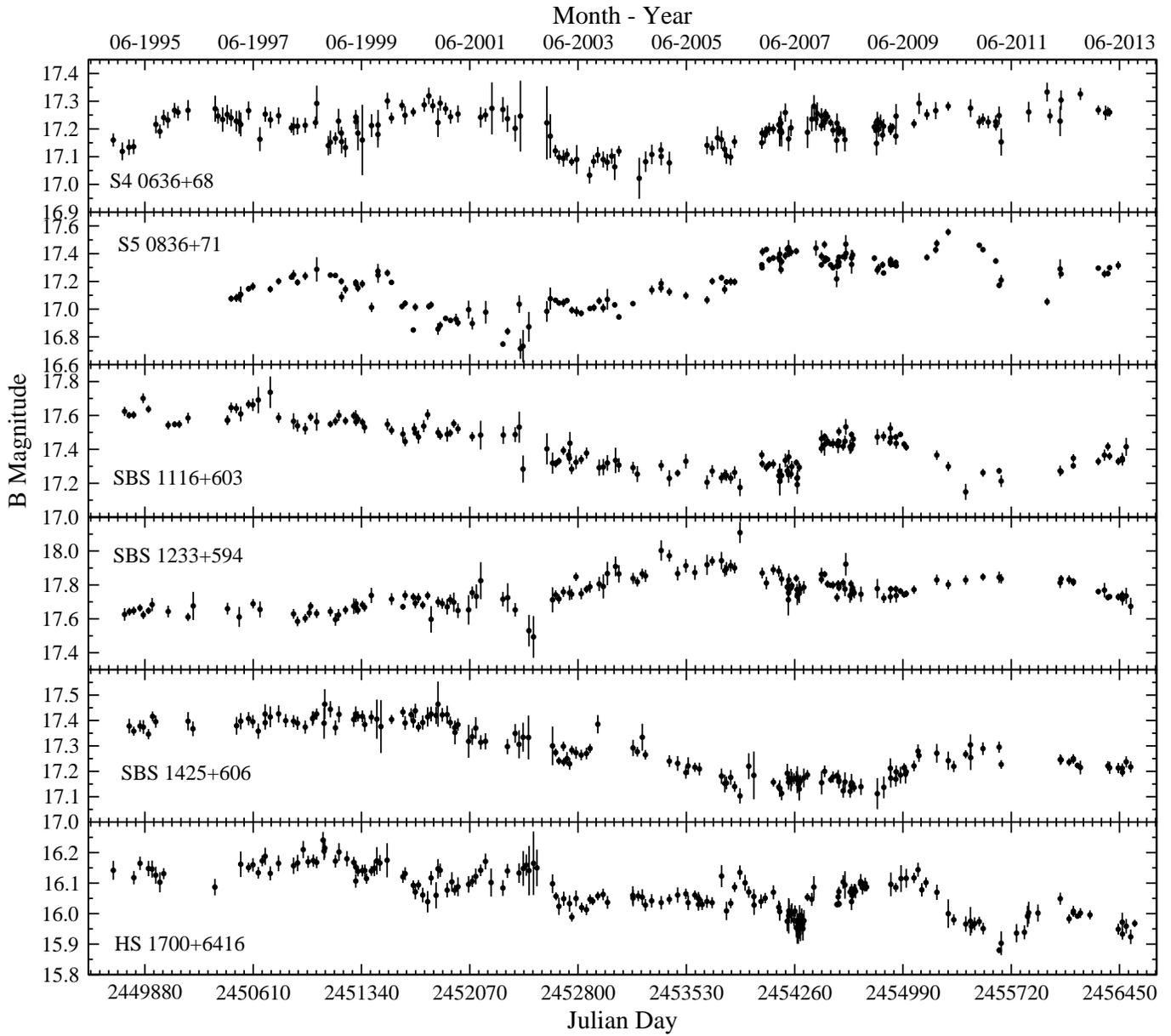

\gridline{\fig{fig13.eps}{1.0\textwidth}{}
          }
\caption{$B$ band light curves of the six quasars with spectroscopic 
observations. The $R$ band light curves of these 
objects were combined with the spectroscopic data as explaned in 
section~\ref{subsec:LC} and there are shown in Figures~\ref{fig:s40636} 
to~\ref{fig:hs1700}.
\label{fig:phot_lc_B1}}
\end{figure*}

\begin{figure*}
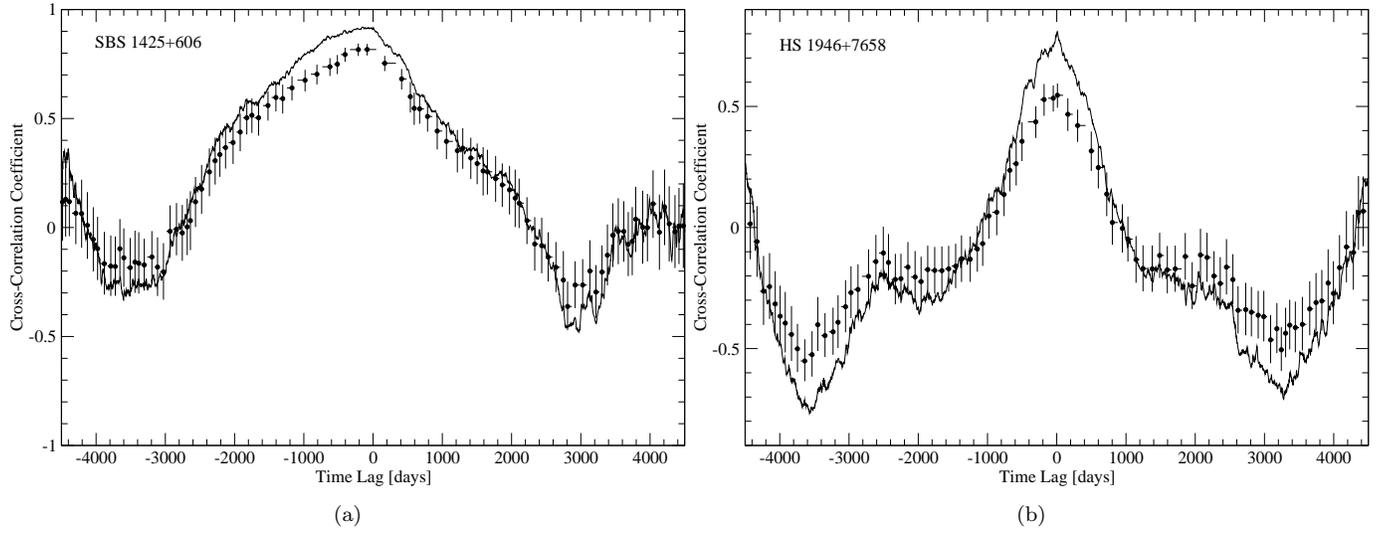

\gridline{\fig{fig14a.eps}{0.5\textwidth}{(a)}
          \fig{fig14b.eps}{0.5\textwidth}{(b)}
          }
\caption{Cross correlation between the $B$ and $R$ bands in (a) SBS\,1425+606 
and (b) HS\,1946+7658. The ICCF is plotted as the solid line while the ZDCF 
is plotted as black points with uncertainties. 
\label{fig:BRccf}}
\end{figure*}

\end{document}